\documentclass[prl,twocolumn,nofootinbib,longbibliography]{revtex4-1}
\usepackage{graphicx}
\usepackage{color}
\usepackage{epsfig}
\usepackage[caption=false]{subfig}
\usepackage{grffile}
\usepackage{amsfonts, amsmath, amsthm, amssymb} 
\usepackage{mathtools}
\usepackage{wrapfig, tikz,url,lipsum,float}
\usepackage{multirow}
\usepackage[normalem]{ulem}

\begin{document}
\title{Geometry-controlled Failure Mechanisms of Amorphous Solids on the Nanoscale}

\author{Kallol Paul$^{1}$}
\author{Ratul Dasgupta$^{2}$} 
\author{J\"urgen Horbach$^{3}$}
\author{Smarajit Karmakar$^{1}$}
\email{smarajit@tifrh.res.in}

\affiliation{
$^1$ TIFR Center for Interdisciplinary Science, Tata Institute of 
Fundamental Research, 36/P Gopanpally Village, Serilingampally Mandal, RR District, Hyderabad, 500075, Telangana, India,\\
$^2$ Department of Chemical Engineering, IIT Bombay, Powai, 
Mumbai 400076, India,\\
$^3$ Institute f\"ur Theoretische Physik II, Heinrich-Heine-Universit\"at D\"usseldorf, 
40225 D\"usseldorf, Germany}

%
%

\keywords{Cavitation $|$ Necking $|$ Ductile $|$ Brittle}

\begin{abstract}
Amorphous solids, confined on the nano-scale, exhibit a wealth of
novel phenomena yet to be explored. In particular, the response of
such solids to a mechanical load is not well understood and, as has
been demonstrated experimentally, it differs strongly from bulk
samples made of the same materials.  Failure patterns and mechanisms
are strongly affected by the geometry of the confinement and the
interplay between interfacial effects in the sample and the time
scale, imposed by an external mechanical field. Here, we present
the mechanism of cavity formation in a confined model glass, subjected
to expansion with a constant strain rate.  This system is studied
for varying geometric aspect ratio and sample size.  Our results
show that for a given temperature and straining condition, the
sample shows cavitation when the aspect ratio reaches a critical
value and below this aspect ratio the sample breaks by forming a neck.
The critical aspect ratio is associated with a critical curvature of
the neck that depends on strain rate and temperature. If this critical 
curvature is exceeded, the free energy of the system is minimized 
by the formation of a cavity. Our study reveals a novel mechanism of
cavity formation on the nanoscale. This is probably a generic mechanism
for material's failure in small confined systems under mechanical load.
\end{abstract}

%
\maketitle

%
\section*{Introduction}
\begin{figure}[!h]
\begin{center}
\hskip -0.5cm
\includegraphics[scale=0.32, angle = 0]{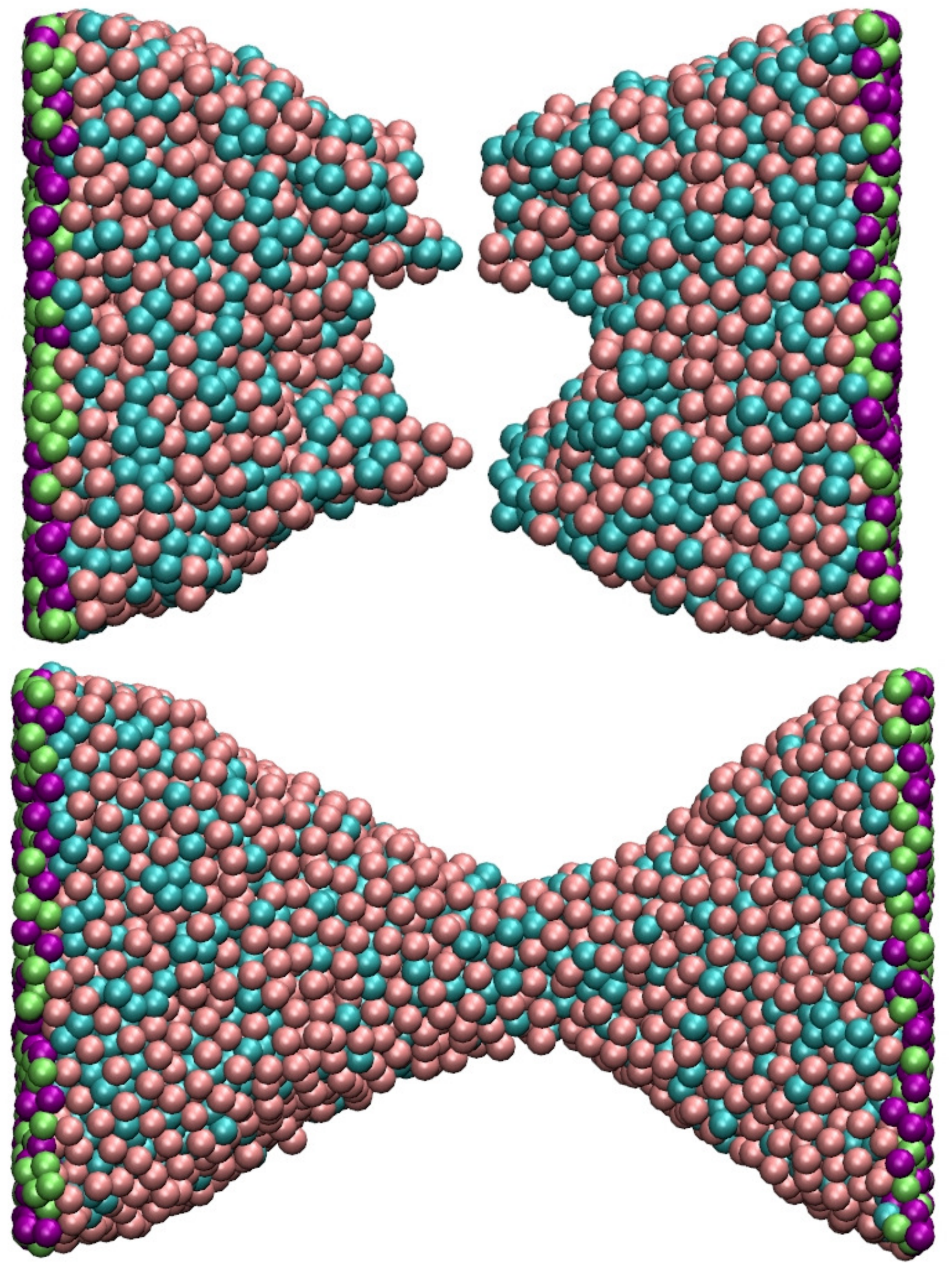}
\caption{Morphologies of Brittle and Ductile failures in model glass forming liquids.}
\label{fig:bittleDuctile}
\end{center}
\end{figure}

Understanding the failure mechanisms of amorphous solids
under external loading is one of the most active research fields
amongst scientific as well as engineering community \cite{ArgonSTZ,Maloney:2006aa,MaloneyLemaitre06,
Hentschel:2010aa,Karmakar:2010ab,Karmakar:2010aa,PhysRevE.84.046105,lagos_das_2016,
HUFNAGEL2016375,doi:10.1146/annurev.matsci.38.060407.130226,ASHBY2006321,RTV11,
article,Shrivastav:2016aa}. This is mainly
due to its importance in fundamental science and industrial
applications \cite{Greer1947, doi:10.1146/annurev.matsci.38.060407.130226}. 
It is experimentally known that amorphous solids have
much larger failure strength than their crystalline counterpart
with similar compositions, but amorphous solids show catastrophic
failure which severely limits their applicability as a useful
design material \cite{Binder2005,PhysRevLett.93.255506,SCHUSTER2007517,
article,doi:10.1063/1.4906305,SUN2015211,HUFNAGEL2016375}. 
Bulk materials that show catastrophic failure by
forming cracks under applied load are generically termed as brittle
materials \cite{PhysRevLett.93.255506,Guo2007,CRETE2014204,doi:10.1111/j.1460-2695.1992.tb00048.x}. 
On the other hand, some materials show significant plastic
deformation before final failure and they are called ductile
materials \cite{PhysRevLett.117.044302,PhysRevB.94.094203}. 
In Fig.~\ref{fig:bittleDuctile}, we show representative
failure morphology for brittleness and ductility for small systems
on the nanoscale. For brittle failure (top panel) the rough failure
surface is noticeable while for ductile failure (bottom panel), the
neck formation is very prominent. Reducing brittleness of strongly
confined amorphous solids is of considerable practical importance.

Recent studies
\cite{doi:10.1063/1.2884584,Chen2013,PhysRevLett.104.215503,Jang2010,
greer2011,kraft2010}
suggest that the nature of plasticity in materials depends very
significantly on their size.  It has been demonstrated experimentally
that failure patterns and mechanisms in nano-scale systems can be very
different from those of bulk samples of the same materials.  Typically,
a bulk metallic glass can sustain only up to 2\% strain before failing via
brittle crack formation. However, uniaxial strain 
experiments, performed in Ref.\cite{PhysRevLett.104.215503} on 
metallic glasses with different 
sample sizes, show that the strain before failure can increase up to 
200\% when the sample size is reduced to 100nm.
For such nano-sized sample, it has been also found that the sample may
break via the formation of necks.  These necks can shrink to chain-like
structures as thin as a few atomic layers. This necking in nano-scale
samples also indicate a ductile rather than a brittle behaviour of the
material in response to deformation.

The change of the mechanical response with reducing
sample size has been also observed in uniaxial compression experiments. In
Ref.~\cite{doi:10.1063/1.2884584}, such experiments on pilers made of
metallic glasses are presented. Here, it is shown that pilers with
diameters larger than 100\,nm show prominent shear band formation,
a hallmark of brittle failure. However, pilers that are less than
100\,nm in diameter display homogeneous plastic deformation without
shear localization.  Similarly, Greer {\it et al.}~\cite{Jang2010} have
demonstrated that size reduction affects the brittle-to-ductile transition
in metallic glasses. They have also indicated that the brittle-to-ductile
transition is associated with a critical size.  Several experiments and
simulations of metallic glasses confirm that size reduction increases
the material ductility due to surface states \cite{PhysRevLett.104.215503,
Chen2013}.
Thus, it is clear that plasticity and failure mechanisms for a material
can change very significantly with the physical size and dimension of
the sample. A microscopic understanding of these observations would be
very important for the systematic development of nano-materials with
specific mechanical properties.

The goal of the present study is to understand the nature of plasticity
when the system size is reduced from bulk to nano-scale and also to find
out the microscopic origin of the observed differences. We have performed
uniaxial elongation simulations of a model glass in nano-confinement
in order to understand the effects of small sample size on the failure
mechanisms and investigate the controlling parameters that determine
whether a material will fail via neck formation or via the formation
of cavities.

\section*{Results}
\begin{figure}[htpb]
\vskip +0.5cm
\begin{center}
\hskip -0.5cm
\includegraphics[scale=0.34, angle = 90]{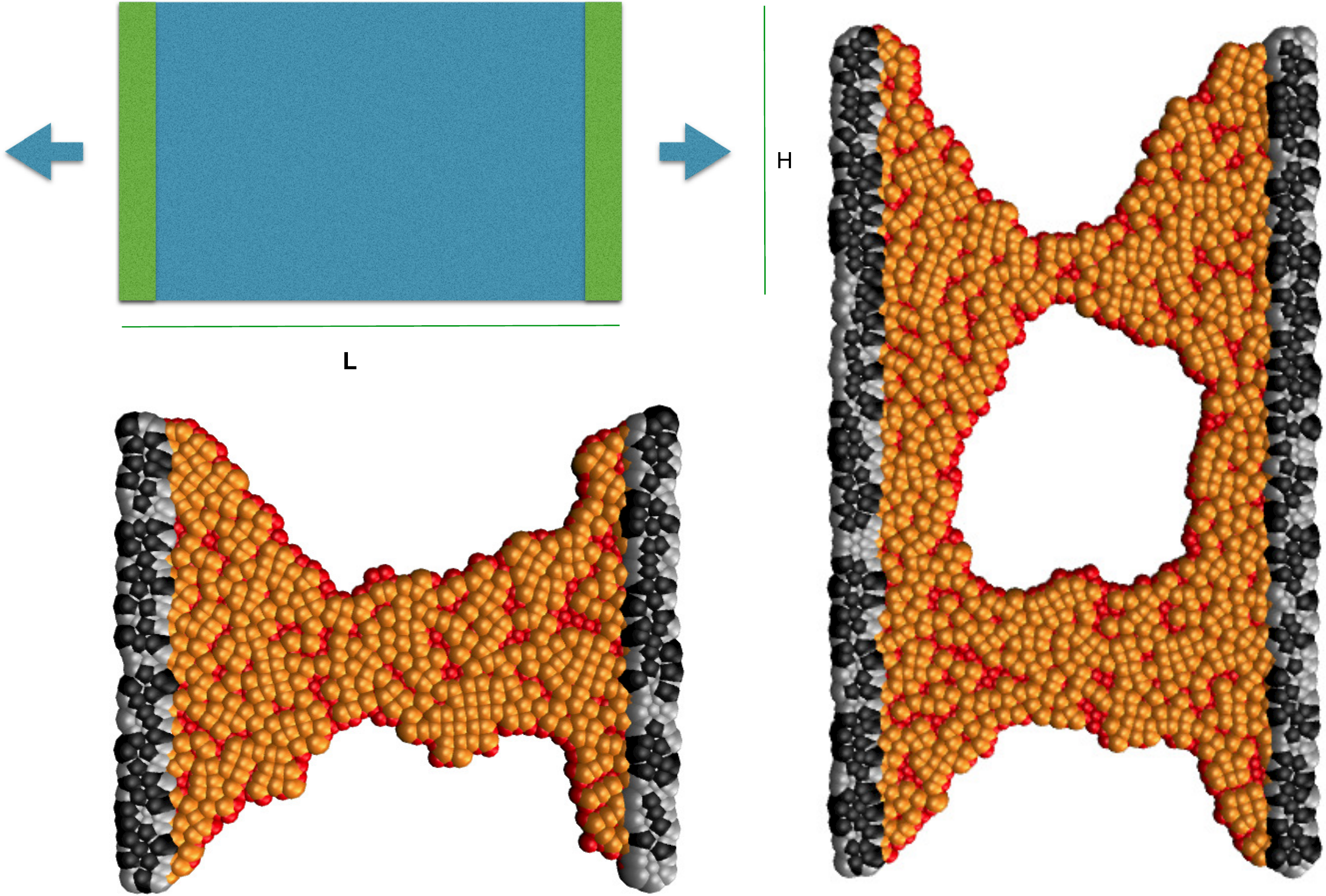}
\caption{Top left panel: Schematic of extensional simulation. Bottom left panel: Representative 
example of failure of a sample via neck formation. Right panel: Similar representative example
of failure of a sample via cavity formation.}
\label{fig:neckCavity}
\end{center}
\end{figure}
We have performed numerical simulations in which we chose a model
glass former (see Supplementary Material (SM) for
further details of the model) and prepared an amorphous solid by
cooling the supercooled liquid at a certain cooling rate (see the
method section for details).  Then, we put a uniaxial load on it
until the material breaks completely.  In all these systems the
numerical experiments are designed as follows.  We first equilibrate
the model liquids at some high temperature and then cool it to a
low temperature below the experimental glass transition temperature
defined as the temperature where the relaxation time reaches a value of \(10^6\).
Then, at that low temperature, we put a barostat and perform constant
pressure and temperature (NPT) simulations at zero pressure such
that one can remove the periodic boundary condition.  We now define
two side walls at the two ends of the solid in the $x$-direction
by freezing the motion of the particles in the two end regions as
depicted in the cartoon in Fig.~\ref{fig:neckCavity}.

\begin{figure}[!htpb]
\begin{center}
\hskip -0.2cm
\includegraphics[scale=0.33, angle = 90]{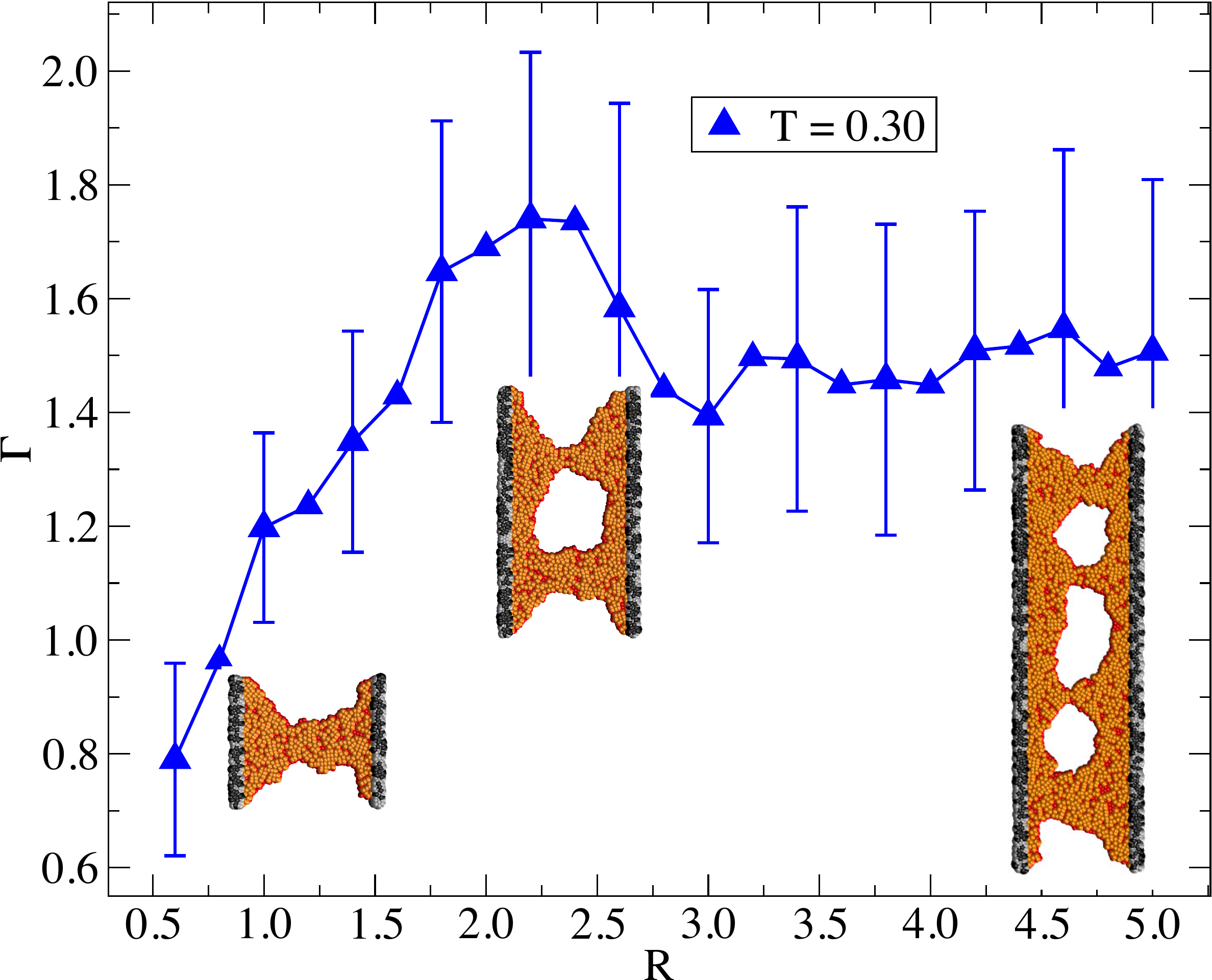}
\caption{Variation of maximum strain, $\Gamma$, with increasing 
aspect ratio, $R$ at $T = 0.30$ and $\dot\gamma = 5\times10^{-6}$. The cross over 
in failure mechanism is also shown by plotting the morphologies of the samples before
the complete failure. }
\label{fig:maxDuctility}
\end{center}
\end{figure}

\begin{figure*}[!htpb]
\begin{center}
\hskip -0.5cm
\includegraphics[scale=0.5,angle=0]{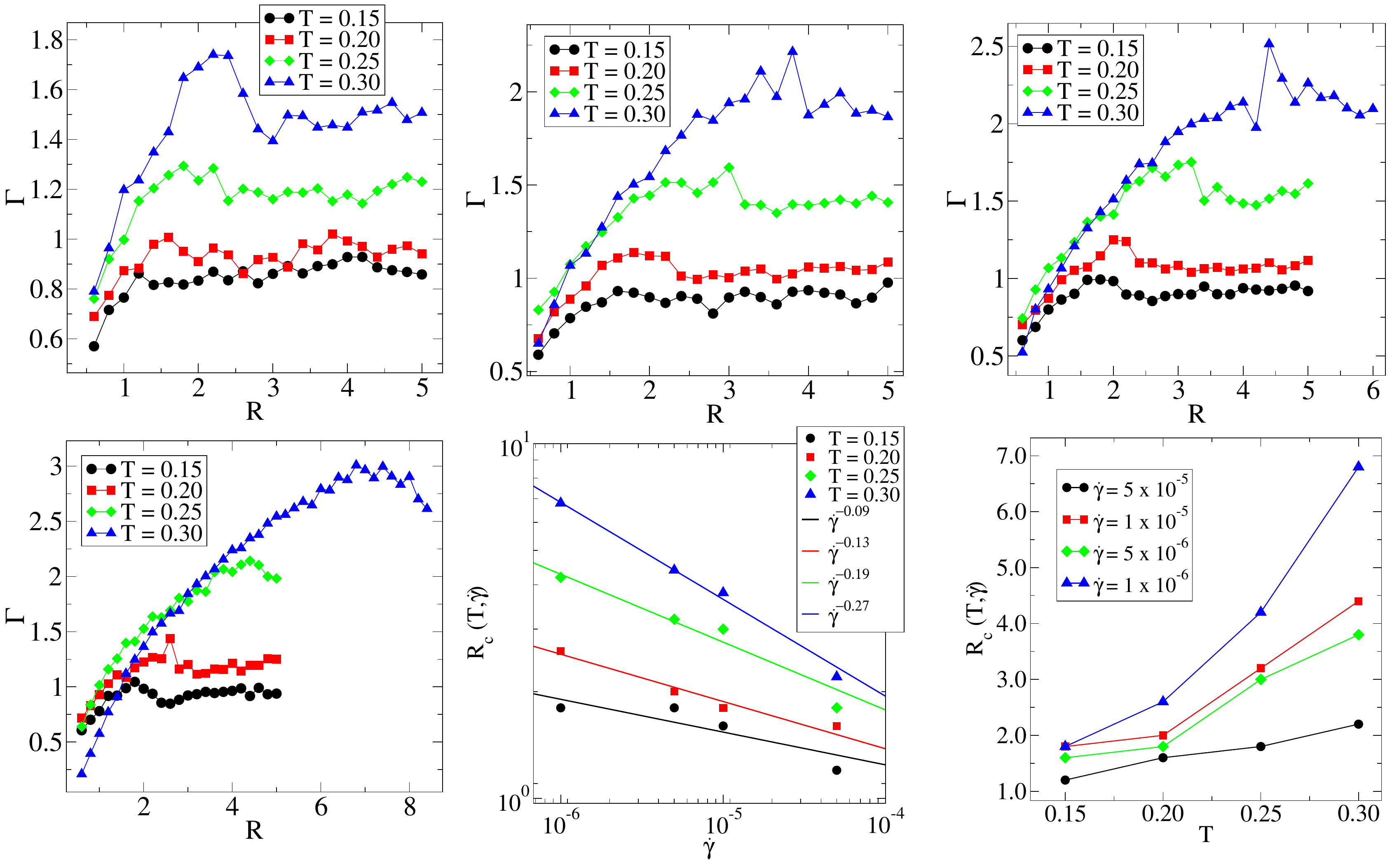}
\caption{Variation of maximum strain, $\Gamma$ as a function of aspect
ratio $R=H/L$ for different temperature $T$ at strain rate 
$\dot{\gamma} = 5\times10^{-5}$ (top-left), $10^{-5}$ 
(top-middle), $5\times10^{-6}$ (top-right), and $(10^{-6}$ 
(bottom-left). Variation of critical aspect ratio, $R_C$ as function of 
strain rate (bottom middle) and temperature (bottom right) (see text for details). }
\label{effectGammadotAndT}
\end{center}
\end{figure*}
The typical size of this wall is around three inter-particle
diameters.  The other boundaries are made free. Next, the walls are
moved by an increment equal in size and opposite in sign, i.e., the
system is subjected to uniaxial strain. To quantify the maximum
deformation that a system can withstand before the failure, we have
defined the maximum strain as
\begin{equation}
\Gamma = \frac{(L_f - L_0)}{L_0},
\end{equation}
where \(L_0\) is the initial (before pulling) length of the system
along the tension direction (\(x\) direction) and
\(L_f\) is the final length when the system breaks into two parts.
We repeat this numerical simulation for different aspect ratio $R
= H/L_0$ ranging from $0.6$ to $5.0$. $H$ is the width of the system
as shown in the schematic diagram in Fig.~\ref{fig:neckCavity}. We
have done extensional simulations for different strain rates and
temperatures keeping $L_0 = 30.0$ for all the systems with different
aspect ratio, $R$.

In Fig.~\ref{fig:maxDuctility}, we have plotted the maximum deformation
$\Gamma$ as a function of aspect ratio, $R$, for $T = 0.30$ at a
strain rate $\dot\gamma = 5\times10^{-6}$. The observation that as
one decreases size of the sample (aspect ratio in this case) the
system starts to show significant change in ductility is in complete
agreement with the experimental results reported in
Refs.~\cite{PhysRevLett.104.215503,doi:10.1063/1.2884584,Jang2010,Chen2013}.
The change in maximum strain is around 40\% with decreasing aspect
ratio at this strain rate. One expects this variation to become
even larger at strain rates that are small and comparable to
experimental strain rates \cite{Sergueeva2004}. As discussed in the
subsequent paragraphs, the variation indeed becomes larger with
decreasing strain rate. Even more surprising results that need
experimental validation is the existence of a critical aspect ratio,
$R_c$, below which one no longer sees an increase in maximum strain,
$\Gamma$. $\Gamma$ actually starts to decrease very rapidly with
decreasing aspect ratio below this critical aspect ratio, $R_c$.
Thus, the $\Gamma$ vs.~$R$ curve shows a peak which signifies the
existence of a favoured geometric aspect ratio at which the material
will show maximum ductility.

Next, we try to characterize the failure patterns as one increases
the aspect ratio, to elucidate the reason for the existence of a
critical aspect ratio and a maximum ductility. As shown in
Fig.~\ref{fig:maxDuctility}, we found that at very small aspect
ratio, $R < R_c$, the system always breaks via neck formation and
the failure mechanisms changes very sharply at the critical aspect
ratio at which one observes the appearance of a single cavity. With
further loading, the cavity increases in size and eventually the
cavity becomes large enough that the system breaks into two fragments.
As one further increases the aspect ratio, the number of cavities
increases by an integer number and one observes the appearance of
multiple cavities. At these aspect ratios, the system eventually
fails by the merger of cavities into a larger cavity in complete
agreement with experimental observations~\cite{DING201589,
MAA201594,An:2011aa,Murali:2011aa,Guan:2013aa, PhysRevLett.117.044302}.
The appearance of cavities above a critical aspect ratio indicates
a ductile-to-brittle transition in the material with increasing
system size. It has been suggested that the change from brittle to
ductile behavior with decreasing system size is related to the
enhanced relaxation on the surface layer
\cite{Jang2010,PhysRevE.84.046105,lagos_das_2016,STEIF1983359,Chen2013,Guo2007}.
To further understand this phenomenon and the effect of surface-induced
relaxation, we have done similar uniaxial elongation simulations
at different strain rate and temperature.

\subsection*{Effect of strain rate $\dot{\gamma}$ and temperature $T$}
In Fig.~\ref{effectGammadotAndT}, we show the maximum strain,
$\Gamma$, as a function of the aspect ratio for different temperatures
at a given strain rate. The critical aspect ratio $R_c$ increases
with increasing temperature and the variation
of $R_c$ with temperature is larger as one decreases the strain
rate, $\dot\gamma$. As shown in Fig.~\ref{effectGammadotAndT}, the
critical aspect ratio changes from $R_c = 1.2$ to $2.2$ for strain
rate $\dot\gamma = 5\times10^{-5}$ (top left panel) as the temperature
increases from $T = 0.15$ to $0.30$, while with slower strain rate
$\dot\gamma = 10^{-6}$ the value of $R_c$ increases from $1.8$ to
around $6.8$ in the same temperature range.  In the bottom middle
panel of Fig.~\ref{effectGammadotAndT}, the critical aspect ratio,
$R_c(T,\dot\gamma)$, is plotted as a function of $\dot\gamma$ for
different temperatures. The critical aspect ratio as a function of
strain rate can be described by a power law, $R_c(T,\dot\gamma)
\sim \dot\gamma^{\eta}$, with exponent $\eta$ varying strongly with
temperature. At low temperature, $\eta$ is very small and approximately
close $-0.1$ and becomes substantially larger as the temperature
is increased towards $T_G$, the calorimetric glass transition
temperature for the model. So at $T = 0.30$, the exponent $\eta
\sim -0.27$ is obtained. In the bottom right panel of the same
figure, the critical aspect ratio is plotted as a function of
temperature for different strain rates. One clearly sees that the
critical aspect ratio where failure happens via cavity formation
increases to larger values, suggesting the expected behaviour of increased
ductility with increasing temperature. This behaviour becomes very
strong as one decreases the strain rate and thus, under experimental
conditions with very slow straining, one can expect a very
strong dependence of the mechanical response on temperature.

\subsection*{Effect of system size}
An increasing aspect ratio while keeping $L_0$ fixed is associated
with an increasing system size. It is not obvious whether the cavity
formation with increasing aspect ratio is solely a geometric effect.
To clarify this issue, we have done simulations with a larger system
size ($N = 10000$) at $R = 1.8$ (smaller than $R_C = 2.0$), $T =
0.2$ and $\dot\gamma = 10^{-5}$, and monitored the failure mechanisms
systematically. We also did another simulation with $N = 500$ at
$R = 3.0$ (bigger than $R_C$). We found that the failure mechanisms
for $R < R_C$ remain neck-like even for larger systems. For $R >
R_C$, it is cavity-dominated even for smaller system size. Within
the studied system size, failure mechanisms seem to be geometric
rather than being a system size effect. Although it can be easily
argued that at system sizes much larger than the one studied in
this work, one will eventually reach the bulk behaviour where the
aspect ratio will \textcolor{green}{no} longer play any role and
the material will behave like a brittle material.  It will be
interesting to find out the crossover system size where one obtains
this behaviour but that system size is expected to be very large
and is beyond the scope of this work.


%
\subsection*{Coalescence of cavities as a generic mechanism for brittle 
failure:}
\begin{figure}[!h]
\begin{center}
\includegraphics[scale=0.3,angle=90]{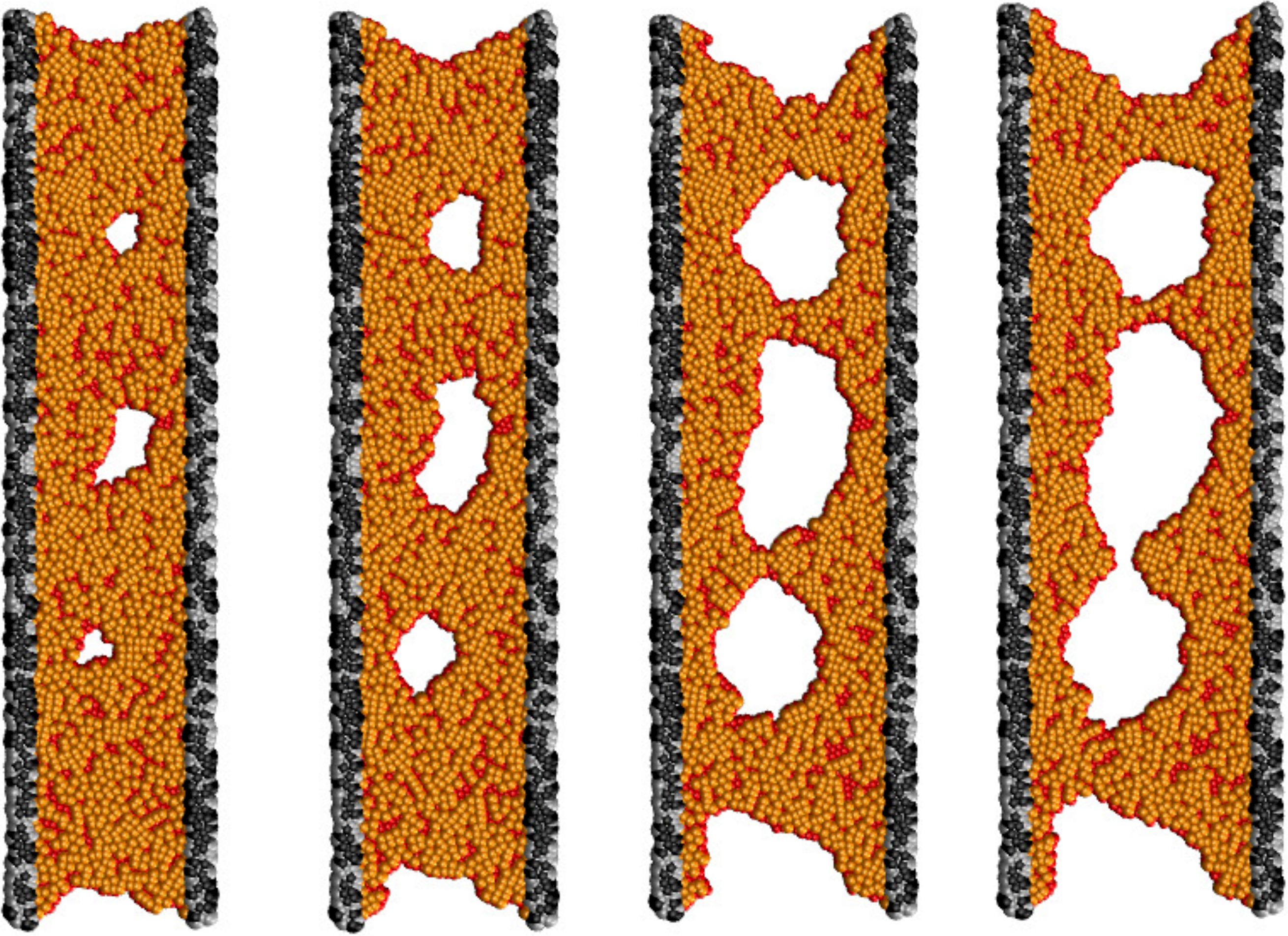}
\caption{The appearance and coalescence of multiple cavities 
for a system with aspect ratio ($R = 4.0$) larger than the critical 
aspect ratio at $T = 0.20$ and strain rate $\dot\gamma = 1\times 10^{-5}$. 
Multiple cavities that are spaced almost equal distance from each other
first appear in the system and then grow with further uniaxial loading.
The cavities then merge to become a large but elongated cavity
before pinching the open boundaries. The rough contour of the fracture
surface can also be seen.}
\label{cavityCoalescence}
\end{center}
\end{figure}
We now focus on the microscopic failure mechanisms for systems with
an aspect ratio much larger than the critical aspect ratio at a
given temperature and strain rate. For systems with an aspect ratio
larger but close to $R_C$, a single cavity appears in the middle
of the system with increasing deformation and then the cavity
increases in size until it meets the open surface to break apart.
With further increase in aspect ratio, one observes the appearance
of two and then three cavities and this continues with increasing
aspect ratio. One also observes that the cavities generally appear
at equal distance from each other and if the straining is stopped
and the cavities are allowed to relax, they remain at that size
without closing them.  This indicates that the cavities formed
during straining are very stable at least in the studied time
duration.  For aspect ratios at which multiple cavities are formed
during straining, first cavities coalesce to become a large single
cavity with further deformation and then finally meets the open
boundary. In Fig.~\ref{cavityCoalescence}, we have shown such a
process of cavity merger for a system with an aspect ratio $R =
4.0$ at $T = 0.20$ and $\dot\gamma = 1\times 10^{-5}$. One can
clearly see that the first three cavities form in the middle of the
sample almost equally spaced. Then, these cavities become larger
in size maintaining roughly their separation. Finally, they merge
to become a larger but elongated cavity before pinching off the
open boundaries.  Coalescence of multiple cavities is found to be
a very generic mechanism for failure in these systems with a larger
aspect ratio. This implies that for bulk brittle materials failure
happens generically via formation and coalescence of cavities. As
for a given aspect ratio, certain numbers of cavities form and then
merge, one should be able to understand the morphology and roughness
of the brittle crack surface from the typical size of these cavities
before their merger. In the rightmost panel of
Fig.~\ref{cavityCoalescence}, one can also see the rough fracture
contour which has a scale roughly equal to the radius of the cavities
before the merger. A much detailed systematic analysis on a larger
system size is needed to have some understanding on this.

\subsection*{Results from 3D systems:}
Until now, results for amorphous solids in two dimensions have been
presented. A natural question is to see to what extent these results
also hold in three dimensional systems. Our initial results of
uniaxial elongation tests for three-dimensional samples under similar
loading conditions and temperatures indeed suggest that the change
in failure mechanism from necking to cavity formation with changing
aspect ratio is also seen in three dimensional systems (see SI for
further details).  So it seems that the existence of a critical
aspect ratio above which cavity formation dominates the failure
mechanism under uniaxial elongation is a generic phenomenon and
does not depend on the dimensionality of the sample. It
will also be very interesting to understand the failure in
three-dimensional nano-sized systems by varying systematically the
cross-sectional area of the sample from square to rectangular to
circular. If one of the side are made much smaller than another
side, one will have a system which will be quasi two dimensional
or like a thin film and failure mechanisms with increasing film
thickness might be important and interesting for practical applications.

\subsection*{Curvature of the open surface}
\begin{figure}[h!]
\begin{center}
{\includegraphics[scale=0.18]{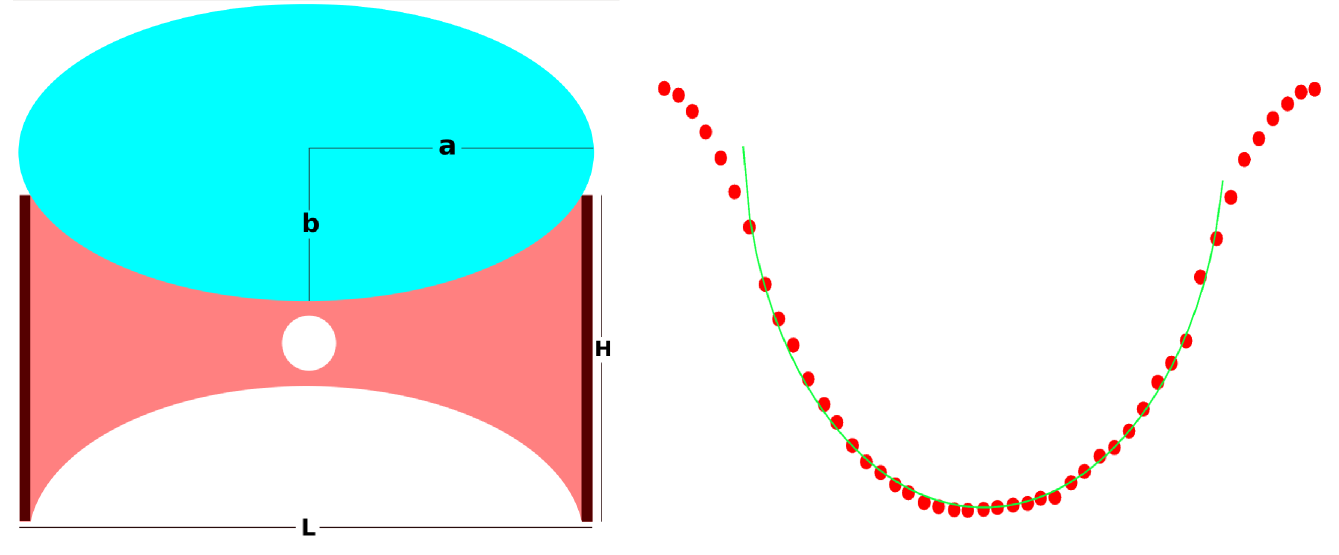}}
\caption{schematic diagram of a suface fitting with an ellipse}
\label{curvatureSchematic}
\end{center}
\end{figure}
To understand the origin of the crossover in the failure mechanism
from necking to cavitation, we now analyze the curvature of the
free surfaces with increasing straining. To this end, we fit the
average morphology of the free surfaces to an ellipse and determine
the radius of curvature ($\mathcal{K} = a^2/b$) at the centre of
the surface (vertex of the minor axis of the ellipse).  In
Fig.~\ref{curvatureSchematic}, we show a typical snapshot of the
surface particles and a fit to it with the equation of an ellipse.
As can be inferred from this snapshot, the fitting of the morphology
of the free surfaces works very well and therefore the estimate of
the curvature is expected to be reliable.

\subsection*{Variation of radius of curvature with aspect ratio} 
\begin{figure}[h!]
\begin{center}
\vskip 0.1in
{\includegraphics[scale=0.35]{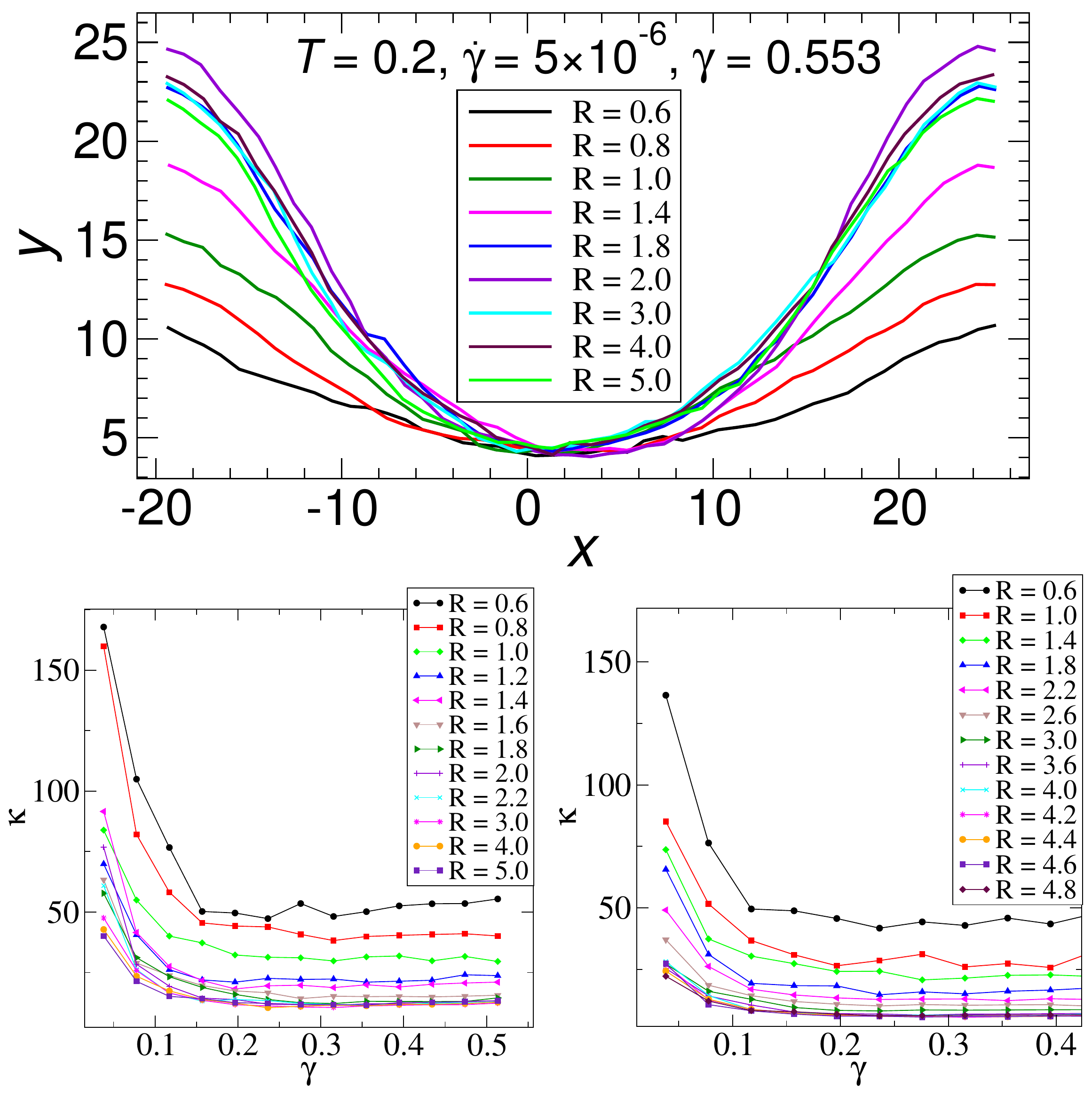}}
\caption{Radius of curvature at a constant strain for different aspect ratio. 
Radius of curvature as a function of strain for different aspect ratio at 
$T = 0.25$ and $\dot{\gamma} = 10^{-5}$ (left), $10^{-6}$ (right)}
\label{curvatureAnalysis}
\end{center}
\end{figure}

In Fig.~\ref{curvatureAnalysis}, we show the curvature of the free
boundary at $T = 0.2$ and $\dot{\gamma} = 5\times 10^{-6}$ for
different aspect ratio.

It is clear that the radius of curvature ($\mathcal{K}$) is decreasing
down to a certain value as the aspect ratio $R$ is increased. The
radius of curvature starts to saturate at $R \simeq 2.0$, which is
close to the critical aspect ratio $R_c$ at $T = 0.2$ and $\dot{\gamma}
= 5\times 10^{-6}$.  This observation clearly indicates that once
$R$ reaches $R_c$, the curvature of the surface can no longer
increase. Thus the system gets constrained to generate more surface
area by increasing the curvature via necking and it leads to cavity
formation in the bulk.  Fig.~\ref{curvatureAnalysis} shows the
evolution of curvature as a function of strain, $\gamma$, for
different aspect ratios $R$ at a particular temperature $T = 0.25$
and strain rates $\dot{\gamma} = 5\times 10^{-5}$ (left panel) and
$10^{-6}$ (right panel).  The radius of curvature initially decreases
monotonically and then reaches a plateau with increasing strain for
each aspect ratio. Beyond the critical aspect
ratio, the curvature seems to become the same for all aspect ratios.
The aspect ratio at which this happens tends to increase with
decreasing strain as can be observed from the data shown in the
right panel of Fig.~\ref{curvatureAnalysis}.

\begin{figure}[h!]
\begin{center}
\vskip -0.1in
{\includegraphics[scale=0.32]{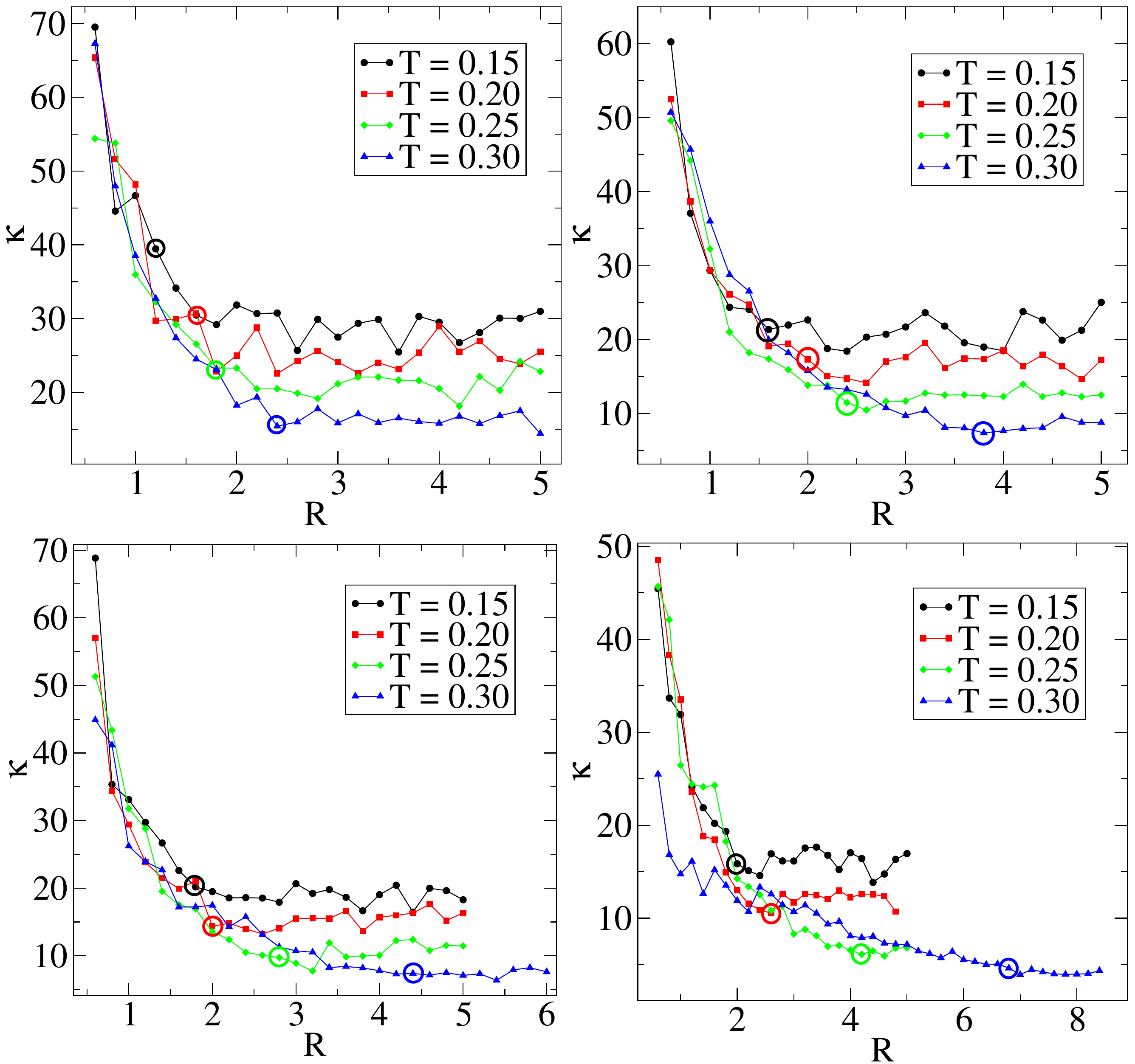}}
\caption{Radius of curvature as a function of aspect ratio for different temperature and strain rate for different temperature $T$ at strain rate $\dot{\gamma} = 5\times10^{-5}$ (top-left), $10^{-5}$ (top-right), $5\times10^{-6}$ (bottom-left), and $(10^{-6}$ (bottom-right)}
\label{curvatureVsR}
\end{center}
\end{figure} 
In Fig.~\ref{curvatureVsR}, we show the radius of curvature
$\mathcal{K}$ at $\gamma = 0.236$ as a function of the aspect ratio
$R$ for different temperatures $T$ and strain rates $\dot{\gamma}$.
For all the studied cases, the radius of curvature starts to saturate
after values of $R$ that are close to the critical aspect ratio,
$R_c$, at that $T$ and $\dot\gamma$. In this figure, we also highlight
the value of the corresponding $R_c$, that we computed earlier from
the variation of $\Gamma$. The correspondence between the value of
$R_c$ at which $\Gamma$ vs.~$R$ has a peak and the value at which
$\mathcal{K}$ vs.~$R$ reaches a saturation value, is undoubtedly
very good (see SI for further details).

\section*{Conclusions}
We have performed uniaxial extensile deformation simulations of a
model amorphous solids to understand the microscopic origin of
cavitation and its connection to ductility in nano-sized amorphous
solids. In this work, we have explored some of the main controlling
parameters that determine whether the solid will
fail via formation of cavities or via necking when subjected to
external uniaxial extensional load. Our findings clearly suggest
that the aspect ratio of the sample is one of the key parameters
that control whether the failure will be via necking or will be
dominated by cavity formation. We also showed for the first time
that there exists a critical aspect ratio at which the ductility
of the material is maximal and the failure mechanism is very different
below and above this critical aspect ratio.  Our results confirm
that all the samples having an aspect ratio below its critical value
will fail via necking whereas samples having aspect ratio larger
than the critical value always forms cavities inside bulk before
it breaks into two segments. Also, the number of cavities increases
as one increases the aspect ratio.  Our observations also show that
this critical aspect ratio is a strong function of temperature and
strain rate as the value of critical aspect ratio becomes larger
at lower strain rate and higher temperature.
Thus the ductility of amorphous solid on the nano-scale depends
strongly on temperature and strain rate. Not much system size
dependence in the failure mechanism is observed within the studied
system size range and it completely depends upon its geometrical
aspect ratio. Thus the conventional notion of enhanced relaxation
due to surface states leading to an increase in ductility of the
material at nano-scale may not be completely correct as geometric
parameters also seem to play a big role. We believe that our
observation of a crossover between neck-like failure to cavity-dominated
failure across a critical aspect ratio will help us in future to
understand why formation and subsequent merger of cavities are
generic mechanisms for brittle failure on small length scales.

\section{Method Section}
We have studied a generic model binary glass former in two dimensions
~\cite{C5SM02200B}.  The equations of motion are integrated with a 
leap-frog algorithm, using an integration time step of $\delta t= 0.005\,\tau$
with $\tau=\sqrt{m \sigma_{\rm AA}^2/\varepsilon_{\rm AA}}$.
The temperature was kept constant via a Berendsen thermostat.
The number of particles \(N\) varied from $648$ to $5400$ and the
bi-dispersity ratio was 65 : 35 (A : B). 

In our simulations, we used a protocol similar to in the recent
study by Babu {\it et al.}~\cite{C5SM02200B}: First, we generate 32
independent configurations at the high temperature $T=1.0$. To this,
the end, simulations over $10^6$ time steps in the $NAT$ ensemble are
performed, applying periodic boundary conditions (PBC) in the two
spatial directions.  Then, these configurations are cooled down to
a desired temperature in the range $0.15\leq T \leq 0.3$ with a 
cooling rate of 0.001, followed in each case by an $NPT$ simulation
at zero pressure, i.e.~$P=0.0$, for another \(2\times10^5\) steps.
Then, we remove the PBC, introducing free boundaries in the $y$ direction
and confining the system in $x$ direction by two disordered walls.
The two walls are obtained by pinning particles in layers of thickness
$3\,\sigma_{\rm AA}$ on both sides in $x$ direction. Now the system
is at the desired temperature and zero pressure, and both walls are
uniaxially pulled in the opposite direction with a constant strain rate
$\dot{\gamma}$.  We vary the strain rate in the range $5\times10^{-5}\leq
\dot{\gamma} \leq 10^{-6}$. At each strain rate, the system is
deformed until it breaks completely into two parts.

\acknowledgements{
We would like to thank Pinaki Chaudhuri, Surajit Sengupta and Srikanth Sastry
for useful discussion. }


\bibliography{cavitation_v5}

\begin{thebibliography}{39}%
\makeatletter
\providecommand \@ifxundefined [1]{%
 \@ifx{#1\undefined}
}%
\providecommand \@ifnum [1]{%
 \ifnum #1\expandafter \@firstoftwo
 \else \expandafter \@secondoftwo
 \fi
}%
\providecommand \@ifx [1]{%
 \ifx #1\expandafter \@firstoftwo
 \else \expandafter \@secondoftwo
 \fi
}%
\providecommand \natexlab [1]{#1}%
\providecommand \enquote  [1]{``#1''}%
\providecommand \bibnamefont  [1]{#1}%
\providecommand \bibfnamefont [1]{#1}%
\providecommand \citenamefont [1]{#1}%
\providecommand \href@noop [0]{\@secondoftwo}%
\providecommand \href [0]{\begingroup \@sanitize@url \@href}%
\providecommand \@href[1]{\@@startlink{#1}\@@href}%
\providecommand \@@href[1]{\endgroup#1\@@endlink}%
\providecommand \@sanitize@url [0]{\catcode `\\12\catcode `\$12\catcode
  `\&12\catcode `\#12\catcode `\^12\catcode `\_12\catcode `\%12\relax}%
\providecommand \@@startlink[1]{}%
\providecommand \@@endlink[0]{}%
\providecommand \url  [0]{\begingroup\@sanitize@url \@url }%
\providecommand \@url [1]{\endgroup\@href {#1}{\urlprefix }}%
\providecommand \urlprefix  [0]{URL }%
\providecommand \Eprint [0]{\href }%
\providecommand \doibase [0]{http://dx.doi.org/}%
\providecommand \selectlanguage [0]{\@gobble}%
\providecommand \bibinfo  [0]{\@secondoftwo}%
\providecommand \bibfield  [0]{\@secondoftwo}%
\providecommand \translation [1]{[#1]}%
\providecommand \BibitemOpen [0]{}%
\providecommand \bibitemStop [0]{}%
\providecommand \bibitemNoStop [0]{.\EOS\space}%
\providecommand \EOS [0]{\spacefactor3000\relax}%
\providecommand \BibitemShut  [1]{\csname bibitem#1\endcsname}%
\let\auto@bib@innerbib\@empty
\bibitem [{\citenamefont {Argon}(1979)}]{ArgonSTZ}%
  \BibitemOpen
  \bibfield  {author} {\bibinfo {author} {\bibfnamefont {A.}~\bibnamefont
  {Argon}},\ }\href {\doibase http://dx.doi.org/10.1016/0001-6160(79)90055-5}
  {\bibfield  {journal} {\bibinfo  {journal} {Acta Metallurgica}\ }\textbf
  {\bibinfo {volume} {27}},\ \bibinfo {pages} {47 } (\bibinfo {year}
  {1979})}\BibitemShut {NoStop}%
\bibitem [{\citenamefont {Maloney}\ and\ \citenamefont
  {Lema{\^\i}tre}(2006)}]{Maloney:2006aa}%
  \BibitemOpen
  \bibfield  {author} {\bibinfo {author} {\bibfnamefont {C.~E.}\ \bibnamefont
  {Maloney}}\ and\ \bibinfo {author} {\bibfnamefont {A.}~\bibnamefont
  {Lema{\^\i}tre}},\ }\href {\doibase 10.1103/PhysRevE.74.016118} {\bibfield
  {journal} {\bibinfo  {journal} {Physical Review E}\ }\textbf {\bibinfo
  {volume} {74}},\ \bibinfo {pages} {016118} (\bibinfo {year}
  {2006})}\BibitemShut {NoStop}%
\bibitem [{\citenamefont {Maloney}\ and\ \citenamefont
  {Lema\^{i}tre}(2006)}]{MaloneyLemaitre06}%
  \BibitemOpen
  \bibfield  {author} {\bibinfo {author} {\bibfnamefont {C.}~\bibnamefont
  {Maloney}}\ and\ \bibinfo {author} {\bibfnamefont {A.}~\bibnamefont
  {Lema\^{i}tre}},\ }\href {\doibase 10.1103/PhysRevE.74.016118} {\bibfield
  {journal} {\bibinfo  {journal} {Phys. Rev. E}\ }\textbf {\bibinfo {volume}
  {74}},\ \bibinfo {pages} {016118} (\bibinfo {year} {2006})}\BibitemShut
  {NoStop}%
\bibitem [{\citenamefont {Hentschel}\ \emph {et~al.}(2010)\citenamefont
  {Hentschel}, \citenamefont {Karmakar}, \citenamefont {Lerner},\ and\
  \citenamefont {Procaccia}}]{Hentschel:2010aa}%
  \BibitemOpen
  \bibfield  {author} {\bibinfo {author} {\bibfnamefont {H.~G.~E.}\
  \bibnamefont {Hentschel}}, \bibinfo {author} {\bibfnamefont {S.}~\bibnamefont
  {Karmakar}}, \bibinfo {author} {\bibfnamefont {E.}~\bibnamefont {Lerner}}, \
  and\ \bibinfo {author} {\bibfnamefont {I.}~\bibnamefont {Procaccia}},\ }\href
  {\doibase 10.1103/PhysRevLett.104.025501} {\bibfield  {journal} {\bibinfo
  {journal} {Physical Review Letters}\ }\textbf {\bibinfo {volume} {104}},\
  \bibinfo {pages} {025501} (\bibinfo {year} {2010})}\BibitemShut {NoStop}%
\bibitem [{\citenamefont {Karmakar}\ \emph
  {et~al.}(2010{\natexlab{a}})\citenamefont {Karmakar}, \citenamefont {Lerner},
  \citenamefont {Procaccia},\ and\ \citenamefont {Zylberg}}]{Karmakar:2010ab}%
  \BibitemOpen
  \bibfield  {author} {\bibinfo {author} {\bibfnamefont {S.}~\bibnamefont
  {Karmakar}}, \bibinfo {author} {\bibfnamefont {E.}~\bibnamefont {Lerner}},
  \bibinfo {author} {\bibfnamefont {I.}~\bibnamefont {Procaccia}}, \ and\
  \bibinfo {author} {\bibfnamefont {J.}~\bibnamefont {Zylberg}},\ }\href
  {\doibase 10.1103/PhysRevE.82.031301} {\bibfield  {journal} {\bibinfo
  {journal} {Physical Review E}\ }\textbf {\bibinfo {volume} {82}},\ \bibinfo
  {pages} {031301} (\bibinfo {year} {2010}{\natexlab{a}})}\BibitemShut
  {NoStop}%
\bibitem [{\citenamefont {Karmakar}\ \emph
  {et~al.}(2010{\natexlab{b}})\citenamefont {Karmakar}, \citenamefont
  {Lerner},\ and\ \citenamefont {Procaccia}}]{Karmakar:2010aa}%
  \BibitemOpen
  \bibfield  {author} {\bibinfo {author} {\bibfnamefont {S.}~\bibnamefont
  {Karmakar}}, \bibinfo {author} {\bibfnamefont {E.}~\bibnamefont {Lerner}}, \
  and\ \bibinfo {author} {\bibfnamefont {I.}~\bibnamefont {Procaccia}},\ }\href
  {\doibase 10.1103/PhysRevE.82.055103} {\bibfield  {journal} {\bibinfo
  {journal} {Physical Review E}\ }\textbf {\bibinfo {volume} {82}},\ \bibinfo
  {pages} {055103} (\bibinfo {year} {2010}{\natexlab{b}})}\BibitemShut
  {NoStop}%
\bibitem [{\citenamefont {Dauchot}\ \emph {et~al.}(2011)\citenamefont
  {Dauchot}, \citenamefont {Karmakar}, \citenamefont {Procaccia},\ and\
  \citenamefont {Zylberg}}]{PhysRevE.84.046105}%
  \BibitemOpen
  \bibfield  {author} {\bibinfo {author} {\bibfnamefont {O.}~\bibnamefont
  {Dauchot}}, \bibinfo {author} {\bibfnamefont {S.}~\bibnamefont {Karmakar}},
  \bibinfo {author} {\bibfnamefont {I.}~\bibnamefont {Procaccia}}, \ and\
  \bibinfo {author} {\bibfnamefont {J.}~\bibnamefont {Zylberg}},\ }\href
  {\doibase 10.1103/PhysRevE.84.046105} {\bibfield  {journal} {\bibinfo
  {journal} {Phys. Rev. E}\ }\textbf {\bibinfo {volume} {84}},\ \bibinfo
  {pages} {046105} (\bibinfo {year} {2011})}\BibitemShut {NoStop}%
\bibitem [{\citenamefont {Lagos}\ and\ \citenamefont
  {Das}(2016)}]{lagos_das_2016}%
  \BibitemOpen
  \bibfield  {author} {\bibinfo {author} {\bibfnamefont {M.}~\bibnamefont
  {Lagos}}\ and\ \bibinfo {author} {\bibfnamefont {R.}~\bibnamefont {Das}},\
  }\href {\doibase 10.4208/aamm.2013.m439} {\bibfield  {journal} {\bibinfo
  {journal} {Advances in Applied Mathematics and Mechanics}\ }\textbf {\bibinfo
  {volume} {8}},\ \bibinfo {pages} {485} (\bibinfo {year} {2016})}\BibitemShut
  {NoStop}%
\bibitem [{\citenamefont {Hufnagel}\ \emph {et~al.}(2016)\citenamefont
  {Hufnagel}, \citenamefont {Schuh},\ and\ \citenamefont
  {Falk}}]{HUFNAGEL2016375}%
  \BibitemOpen
  \bibfield  {author} {\bibinfo {author} {\bibfnamefont {T.~C.}\ \bibnamefont
  {Hufnagel}}, \bibinfo {author} {\bibfnamefont {C.~A.}\ \bibnamefont {Schuh}},
  \ and\ \bibinfo {author} {\bibfnamefont {M.~L.}\ \bibnamefont {Falk}},\
  }\href {\doibase https://doi.org/10.1016/j.actamat.2016.01.049} {\bibfield
  {journal} {\bibinfo  {journal} {Acta Materialia}\ }\textbf {\bibinfo {volume}
  {109}},\ \bibinfo {pages} {375 } (\bibinfo {year} {2016})}\BibitemShut
  {NoStop}%
\bibitem [{\citenamefont
  {Chen}(2008)}]{doi:10.1146/annurev.matsci.38.060407.130226}%
  \BibitemOpen
  \bibfield  {author} {\bibinfo {author} {\bibfnamefont {M.}~\bibnamefont
  {Chen}},\ }\href {\doibase 10.1146/annurev.matsci.38.060407.130226}
  {\bibfield  {journal} {\bibinfo  {journal} {Annual Review of Materials
  Research}\ }\textbf {\bibinfo {volume} {38}},\ \bibinfo {pages} {445}
  (\bibinfo {year} {2008})},\ \Eprint
  {http://arxiv.org/abs/https://doi.org/10.1146/annurev.matsci.38.060407.130226}
  {https://doi.org/10.1146/annurev.matsci.38.060407.130226} \BibitemShut
  {NoStop}%
\bibitem [{\citenamefont {Ashby}\ and\ \citenamefont
  {Greer}(2006)}]{ASHBY2006321}%
  \BibitemOpen
  \bibfield  {author} {\bibinfo {author} {\bibfnamefont {M.}~\bibnamefont
  {Ashby}}\ and\ \bibinfo {author} {\bibfnamefont {A.}~\bibnamefont {Greer}},\
  }\href {\doibase https://doi.org/10.1016/j.scriptamat.2005.09.051} {\bibfield
   {journal} {\bibinfo  {journal} {Scripta Materialia}\ }\textbf {\bibinfo
  {volume} {54}},\ \bibinfo {pages} {321 } (\bibinfo {year} {2006})},\ \bibinfo
  {note} {viewpoint set no: 37. On mechanical behavior of metallic
  glasses}\BibitemShut {NoStop}%
\bibitem [{\citenamefont {Rodney}\ \emph {et~al.}(2011)\citenamefont {Rodney},
  \citenamefont {Tanguy},\ and\ \citenamefont {Vandembroucq}}]{RTV11}%
  \BibitemOpen
  \bibfield  {author} {\bibinfo {author} {\bibfnamefont {D.}~\bibnamefont
  {Rodney}}, \bibinfo {author} {\bibfnamefont {A.}~\bibnamefont {Tanguy}}, \
  and\ \bibinfo {author} {\bibfnamefont {D.}~\bibnamefont {Vandembroucq}},\
  }\href@noop {} {\bibfield  {journal} {\bibinfo  {journal} {Modelling and
  Simulation in Materials Science and Engineering}\ }\textbf {\bibinfo {volume}
  {19}},\ \bibinfo {pages} {083001} (\bibinfo {year} {2011})}\BibitemShut
  {NoStop}%
\bibitem [{\citenamefont {Schuh}\ \emph {et~al.}(2007)\citenamefont {Schuh},
  \citenamefont {Hufnagel},\ and\ \citenamefont {Ramamurty}}]{article}%
  \BibitemOpen
  \bibfield  {author} {\bibinfo {author} {\bibfnamefont {C.}~\bibnamefont
  {Schuh}}, \bibinfo {author} {\bibfnamefont {T.}~\bibnamefont {Hufnagel}}, \
  and\ \bibinfo {author} {\bibfnamefont {U.}~\bibnamefont {Ramamurty}},\ }\href
  {\doibase 10.1016/j.actamat.2007.01.052} {\bibfield  {journal} {\bibinfo
  {journal} {Acta Materialia - ACTA MATER}\ }\textbf {\bibinfo {volume} {55}},\
  \bibinfo {pages} {4067} (\bibinfo {year} {2007})}\BibitemShut {NoStop}%
\bibitem [{\citenamefont {Shrivastav}\ \emph {et~al.}(2016)\citenamefont
  {Shrivastav}, \citenamefont {Chaudhuri},\ and\ \citenamefont
  {Horbach}}]{Shrivastav:2016aa}%
  \BibitemOpen
  \bibfield  {author} {\bibinfo {author} {\bibfnamefont {G.~P.}\ \bibnamefont
  {Shrivastav}}, \bibinfo {author} {\bibfnamefont {P.}~\bibnamefont
  {Chaudhuri}}, \ and\ \bibinfo {author} {\bibfnamefont {J.}~\bibnamefont
  {Horbach}},\ }\href {\doibase 10.1103/PhysRevE.94.042605} {\bibfield
  {journal} {\bibinfo  {journal} {Physical Review E}\ }\textbf {\bibinfo
  {volume} {94}},\ \bibinfo {pages} {042605} (\bibinfo {year}
  {2016})}\BibitemShut {NoStop}%
\bibitem [{\citenamefont {Greer}(1995)}]{Greer1947}%
  \BibitemOpen
  \bibfield  {author} {\bibinfo {author} {\bibfnamefont {A.~L.}\ \bibnamefont
  {Greer}},\ }\href {\doibase 10.1126/science.267.5206.1947} {\bibfield
  {journal} {\bibinfo  {journal} {Science}\ }\textbf {\bibinfo {volume}
  {267}},\ \bibinfo {pages} {1947} (\bibinfo {year} {1995})}\BibitemShut
  {NoStop}%
\bibitem [{\citenamefont {Binder}\ \emph {et~al.}(2005)\citenamefont {Binder},
  \citenamefont {Horbach}, \citenamefont {Kob},\ and\ \citenamefont
  {Winkler}}]{Binder2005}%
  \BibitemOpen
  \bibfield  {author} {\bibinfo {author} {\bibfnamefont {K.}~\bibnamefont
  {Binder}}, \bibinfo {author} {\bibfnamefont {J.}~\bibnamefont {Horbach}},
  \bibinfo {author} {\bibfnamefont {W.}~\bibnamefont {Kob}}, \ and\ \bibinfo
  {author} {\bibfnamefont {A.}~\bibnamefont {Winkler}},\ }\enquote {\bibinfo
  {title} {Computer simulation of molten and glassy silica and its mixtures
  with sodium oxide and aluminium oxide},}\ in\ \href {\doibase
  10.1007/0-387-25953-8_4} {\emph {\bibinfo {booktitle} {Complex Inorganic
  Solids: Structural, Stability, and Magnetic Properties of Alloys}}},\
  \bibinfo {editor} {edited by\ \bibinfo {editor} {\bibfnamefont {P.~E.~A.}\
  \bibnamefont {Turchi}}, \bibinfo {editor} {\bibfnamefont {A.}~\bibnamefont
  {Gonis}}, \bibinfo {editor} {\bibfnamefont {K.}~\bibnamefont {Rajan}}, \ and\
  \bibinfo {editor} {\bibfnamefont {A.}~\bibnamefont {Meike}}}\ (\bibinfo
  {publisher} {Springer US},\ \bibinfo {address} {Boston, MA},\ \bibinfo {year}
  {2005})\ pp.\ \bibinfo {pages} {35--53}\BibitemShut {NoStop}%
\bibitem [{\citenamefont {Schroers}\ and\ \citenamefont
  {Johnson}(2004)}]{PhysRevLett.93.255506}%
  \BibitemOpen
  \bibfield  {author} {\bibinfo {author} {\bibfnamefont {J.}~\bibnamefont
  {Schroers}}\ and\ \bibinfo {author} {\bibfnamefont {W.~L.}\ \bibnamefont
  {Johnson}},\ }\href {\doibase 10.1103/PhysRevLett.93.255506} {\bibfield
  {journal} {\bibinfo  {journal} {Phys. Rev. Lett.}\ }\textbf {\bibinfo
  {volume} {93}},\ \bibinfo {pages} {255506} (\bibinfo {year}
  {2004})}\BibitemShut {NoStop}%
\bibitem [{\citenamefont {Schuster}\ \emph {et~al.}(2007)\citenamefont
  {Schuster}, \citenamefont {Wei}, \citenamefont {Ervin}, \citenamefont
  {Hruszkewycz}, \citenamefont {Miller}, \citenamefont {Hufnagel},\ and\
  \citenamefont {Ramesh}}]{SCHUSTER2007517}%
  \BibitemOpen
  \bibfield  {author} {\bibinfo {author} {\bibfnamefont {B.}~\bibnamefont
  {Schuster}}, \bibinfo {author} {\bibfnamefont {Q.}~\bibnamefont {Wei}},
  \bibinfo {author} {\bibfnamefont {M.}~\bibnamefont {Ervin}}, \bibinfo
  {author} {\bibfnamefont {S.}~\bibnamefont {Hruszkewycz}}, \bibinfo {author}
  {\bibfnamefont {M.}~\bibnamefont {Miller}}, \bibinfo {author} {\bibfnamefont
  {T.}~\bibnamefont {Hufnagel}}, \ and\ \bibinfo {author} {\bibfnamefont
  {K.}~\bibnamefont {Ramesh}},\ }\href {\doibase
  https://doi.org/10.1016/j.scriptamat.2007.05.025} {\bibfield  {journal}
  {\bibinfo  {journal} {Scripta Materialia}\ }\textbf {\bibinfo {volume}
  {57}},\ \bibinfo {pages} {517 } (\bibinfo {year} {2007})}\BibitemShut
  {NoStop}%
\bibitem [{\citenamefont {Scudino}\ \emph {et~al.}(2015)\citenamefont
  {Scudino}, \citenamefont {Shakur~Shahabi}, \citenamefont {Stoica},
  \citenamefont {Kaban}, \citenamefont {Escher}, \citenamefont {K{\"u}hn},
  \citenamefont {Vaughan},\ and\ \citenamefont
  {Eckert}}]{doi:10.1063/1.4906305}%
  \BibitemOpen
  \bibfield  {author} {\bibinfo {author} {\bibfnamefont {S.}~\bibnamefont
  {Scudino}}, \bibinfo {author} {\bibfnamefont {H.}~\bibnamefont
  {Shakur~Shahabi}}, \bibinfo {author} {\bibfnamefont {M.}~\bibnamefont
  {Stoica}}, \bibinfo {author} {\bibfnamefont {I.}~\bibnamefont {Kaban}},
  \bibinfo {author} {\bibfnamefont {B.}~\bibnamefont {Escher}}, \bibinfo
  {author} {\bibfnamefont {U.}~\bibnamefont {K{\"u}hn}}, \bibinfo {author}
  {\bibfnamefont {G.~B.~M.}\ \bibnamefont {Vaughan}}, \ and\ \bibinfo {author}
  {\bibfnamefont {J.}~\bibnamefont {Eckert}},\ }\href {\doibase
  10.1063/1.4906305} {\bibfield  {journal} {\bibinfo  {journal} {Applied
  Physics Letters}\ }\textbf {\bibinfo {volume} {106}},\ \bibinfo {pages}
  {031903} (\bibinfo {year} {2015})},\ \Eprint
  {http://arxiv.org/abs/https://doi.org/10.1063/1.4906305}
  {https://doi.org/10.1063/1.4906305} \BibitemShut {NoStop}%
\bibitem [{\citenamefont {Sun}\ and\ \citenamefont {Wang}(2015)}]{SUN2015211}%
  \BibitemOpen
  \bibfield  {author} {\bibinfo {author} {\bibfnamefont {B.}~\bibnamefont
  {Sun}}\ and\ \bibinfo {author} {\bibfnamefont {W.}~\bibnamefont {Wang}},\
  }\href {\doibase https://doi.org/10.1016/j.pmatsci.2015.05.002} {\bibfield
  {journal} {\bibinfo  {journal} {Progress in Materials Science}\ }\textbf
  {\bibinfo {volume} {74}},\ \bibinfo {pages} {211 } (\bibinfo {year}
  {2015})}\BibitemShut {NoStop}%
\bibitem [{\citenamefont {Guo}\ \emph {et~al.}(2007)\citenamefont {Guo},
  \citenamefont {Yan}, \citenamefont {Wang}, \citenamefont {Tan}, \citenamefont
  {Zhang}, \citenamefont {Sui},\ and\ \citenamefont {Ma}}]{Guo2007}%
  \BibitemOpen
  \bibfield  {author} {\bibinfo {author} {\bibfnamefont {H.}~\bibnamefont
  {Guo}}, \bibinfo {author} {\bibfnamefont {P.~F.}\ \bibnamefont {Yan}},
  \bibinfo {author} {\bibfnamefont {Y.~B.}\ \bibnamefont {Wang}}, \bibinfo
  {author} {\bibfnamefont {J.}~\bibnamefont {Tan}}, \bibinfo {author}
  {\bibfnamefont {Z.~F.}\ \bibnamefont {Zhang}}, \bibinfo {author}
  {\bibfnamefont {M.~L.}\ \bibnamefont {Sui}}, \ and\ \bibinfo {author}
  {\bibfnamefont {E.}~\bibnamefont {Ma}},\ }\href
  {https://doi.org/10.1038/nmat1984} {\bibfield  {journal} {\bibinfo  {journal}
  {Nature Materials}\ }\textbf {\bibinfo {volume} {6}},\ \bibinfo {pages} {735
  EP } (\bibinfo {year} {2007})}\BibitemShut {NoStop}%
\bibitem [{\citenamefont {Cr{\'e}t{\'e}}\ \emph {et~al.}(2014)\citenamefont
  {Cr{\'e}t{\'e}}, \citenamefont {Long{\`e}re},\ and\ \citenamefont
  {Cadou}}]{CRETE2014204}%
  \BibitemOpen
  \bibfield  {author} {\bibinfo {author} {\bibfnamefont {J.}~\bibnamefont
  {Cr{\'e}t{\'e}}}, \bibinfo {author} {\bibfnamefont {P.}~\bibnamefont
  {Long{\`e}re}}, \ and\ \bibinfo {author} {\bibfnamefont {J.}~\bibnamefont
  {Cadou}},\ }\href {\doibase https://doi.org/10.1016/j.cma.2014.03.007}
  {\bibfield  {journal} {\bibinfo  {journal} {Computer Methods in Applied
  Mechanics and Engineering}\ }\textbf {\bibinfo {volume} {275}},\ \bibinfo
  {pages} {204 } (\bibinfo {year} {2014})}\BibitemShut {NoStop}%
\bibitem [{\citenamefont {Li}\ \emph {et~al.}(1992)\citenamefont {Li},
  \citenamefont {Liu}, \citenamefont {Du}, \citenamefont {Hong},\ and\
  \citenamefont {Zhang}}]{doi:10.1111/j.1460-2695.1992.tb00048.x}%
  \BibitemOpen
  \bibfield  {author} {\bibinfo {author} {\bibfnamefont {G.~C.}\ \bibnamefont
  {Li}}, \bibinfo {author} {\bibfnamefont {H.~Q.}\ \bibnamefont {Liu}},
  \bibinfo {author} {\bibfnamefont {M.~L.}\ \bibnamefont {Du}}, \bibinfo
  {author} {\bibfnamefont {Y.~S.}\ \bibnamefont {Hong}}, \ and\ \bibinfo
  {author} {\bibfnamefont {X.}~\bibnamefont {Zhang}},\ }\href {\doibase
  10.1111/j.1460-2695.1992.tb00048.x} {\bibfield  {journal} {\bibinfo
  {journal} {Fatigue \& Fracture of Engineering Materials \& Structures}\
  }\textbf {\bibinfo {volume} {15}},\ \bibinfo {pages} {187} (\bibinfo {year}
  {1992})},\ \Eprint
  {http://arxiv.org/abs/https://onlinelibrary.wiley.com/doi/pdf/10.1111/j.1460-2695.1992.tb00048.x}
  {https://onlinelibrary.wiley.com/doi/pdf/10.1111/j.1460-2695.1992.tb00048.x}
  \BibitemShut {NoStop}%
\bibitem [{\citenamefont {Singh}\ \emph {et~al.}(2016)\citenamefont {Singh},
  \citenamefont {Narasimhan},\ and\ \citenamefont
  {Ramamurty}}]{PhysRevLett.117.044302}%
  \BibitemOpen
  \bibfield  {author} {\bibinfo {author} {\bibfnamefont {I.}~\bibnamefont
  {Singh}}, \bibinfo {author} {\bibfnamefont {R.}~\bibnamefont {Narasimhan}}, \
  and\ \bibinfo {author} {\bibfnamefont {U.}~\bibnamefont {Ramamurty}},\ }\href
  {\doibase 10.1103/PhysRevLett.117.044302} {\bibfield  {journal} {\bibinfo
  {journal} {Phys. Rev. Lett.}\ }\textbf {\bibinfo {volume} {117}},\ \bibinfo
  {pages} {044302} (\bibinfo {year} {2016})}\BibitemShut {NoStop}%
\bibitem [{\citenamefont {Chaudhuri}\ and\ \citenamefont
  {Horbach}(2016)}]{PhysRevB.94.094203}%
  \BibitemOpen
  \bibfield  {author} {\bibinfo {author} {\bibfnamefont {P.}~\bibnamefont
  {Chaudhuri}}\ and\ \bibinfo {author} {\bibfnamefont {J.}~\bibnamefont
  {Horbach}},\ }\href {\doibase 10.1103/PhysRevB.94.094203} {\bibfield
  {journal} {\bibinfo  {journal} {Phys. Rev. B}\ }\textbf {\bibinfo {volume}
  {94}},\ \bibinfo {pages} {094203} (\bibinfo {year} {2016})}\BibitemShut
  {NoStop}%
\bibitem [{\citenamefont {Volkert}\ \emph {et~al.}(2008)\citenamefont
  {Volkert}, \citenamefont {Donohue},\ and\ \citenamefont
  {Spaepen}}]{doi:10.1063/1.2884584}%
  \BibitemOpen
  \bibfield  {author} {\bibinfo {author} {\bibfnamefont {C.~A.}\ \bibnamefont
  {Volkert}}, \bibinfo {author} {\bibfnamefont {A.}~\bibnamefont {Donohue}}, \
  and\ \bibinfo {author} {\bibfnamefont {F.}~\bibnamefont {Spaepen}},\ }\href
  {\doibase 10.1063/1.2884584} {\bibfield  {journal} {\bibinfo  {journal}
  {Journal of Applied Physics}\ }\textbf {\bibinfo {volume} {103}},\ \bibinfo
  {pages} {083539} (\bibinfo {year} {2008})},\ \Eprint
  {http://arxiv.org/abs/https://doi.org/10.1063/1.2884584}
  {https://doi.org/10.1063/1.2884584} \BibitemShut {NoStop}%
\bibitem [{\citenamefont {Chen}\ \emph {et~al.}(2013)\citenamefont {Chen},
  \citenamefont {Jang}, \citenamefont {Guan}, \citenamefont {An}, \citenamefont
  {Goddard},\ and\ \citenamefont {Greer}}]{Chen2013}%
  \BibitemOpen
  \bibfield  {author} {\bibinfo {author} {\bibfnamefont {D.~Z.}\ \bibnamefont
  {Chen}}, \bibinfo {author} {\bibfnamefont {D.}~\bibnamefont {Jang}}, \bibinfo
  {author} {\bibfnamefont {K.~M.}\ \bibnamefont {Guan}}, \bibinfo {author}
  {\bibfnamefont {Q.}~\bibnamefont {An}}, \bibinfo {author} {\bibfnamefont
  {W.~A.}\ \bibnamefont {Goddard}}, \ and\ \bibinfo {author} {\bibfnamefont
  {J.~R.}\ \bibnamefont {Greer}},\ }\href {\doibase 10.1021/nl402384r}
  {\bibfield  {journal} {\bibinfo  {journal} {Nano Letters}\ }\textbf {\bibinfo
  {volume} {13}},\ \bibinfo {pages} {4462} (\bibinfo {year}
  {2013})}\BibitemShut {NoStop}%
\bibitem [{\citenamefont {Luo}\ \emph {et~al.}(2010)\citenamefont {Luo},
  \citenamefont {Wu}, \citenamefont {Huang}, \citenamefont {Wang},\ and\
  \citenamefont {Mao}}]{PhysRevLett.104.215503}%
  \BibitemOpen
  \bibfield  {author} {\bibinfo {author} {\bibfnamefont {J.~H.}\ \bibnamefont
  {Luo}}, \bibinfo {author} {\bibfnamefont {F.~F.}\ \bibnamefont {Wu}},
  \bibinfo {author} {\bibfnamefont {J.~Y.}\ \bibnamefont {Huang}}, \bibinfo
  {author} {\bibfnamefont {J.~Q.}\ \bibnamefont {Wang}}, \ and\ \bibinfo
  {author} {\bibfnamefont {S.~X.}\ \bibnamefont {Mao}},\ }\href {\doibase
  10.1103/PhysRevLett.104.215503} {\bibfield  {journal} {\bibinfo  {journal}
  {Phys. Rev. Lett.}\ }\textbf {\bibinfo {volume} {104}},\ \bibinfo {pages}
  {215503} (\bibinfo {year} {2010})}\BibitemShut {NoStop}%
\bibitem [{\citenamefont {Jang}\ and\ \citenamefont {Greer}(2010)}]{Jang2010}%
  \BibitemOpen
  \bibfield  {author} {\bibinfo {author} {\bibfnamefont {D.}~\bibnamefont
  {Jang}}\ and\ \bibinfo {author} {\bibfnamefont {J.~R.}\ \bibnamefont
  {Greer}},\ }\href {https://doi.org/10.1038/nmat2622} {\bibfield  {journal}
  {\bibinfo  {journal} {Nature Materials}\ }\textbf {\bibinfo {volume} {9}},\
  \bibinfo {pages} {215 EP } (\bibinfo {year} {2010})}\BibitemShut {NoStop}%
\bibitem [{\citenamefont {Greer}\ and\ \citenamefont
  {De~Hosson}(2011)}]{greer2011}%
  \BibitemOpen
  \bibfield  {author} {\bibinfo {author} {\bibfnamefont {J.~R.}\ \bibnamefont
  {Greer}}\ and\ \bibinfo {author} {\bibfnamefont {J.~T.~M.}\ \bibnamefont
  {De~Hosson}},\ }\bibfield  {booktitle} {\emph {\bibinfo {booktitle}
  {Festschrift Vaclav Vitek}},\ }\href {\doibase
  https://doi.org/10.1016/j.pmatsci.2011.01.005} {\bibfield  {journal}
  {\bibinfo  {journal} {Progress in Materials Science}\ }\textbf {\bibinfo
  {volume} {56}},\ \bibinfo {pages} {654} (\bibinfo {year} {2011})}\BibitemShut
  {NoStop}%
\bibitem [{\citenamefont {Kraft}\ \emph {et~al.}(2010)\citenamefont {Kraft},
  \citenamefont {Gruber}, \citenamefont {M{\"o}nig},\ and\ \citenamefont
  {Weygand}}]{kraft2010}%
  \BibitemOpen
  \bibfield  {author} {\bibinfo {author} {\bibfnamefont {O.}~\bibnamefont
  {Kraft}}, \bibinfo {author} {\bibfnamefont {P.~A.}\ \bibnamefont {Gruber}},
  \bibinfo {author} {\bibfnamefont {R.}~\bibnamefont {M{\"o}nig}}, \ and\
  \bibinfo {author} {\bibfnamefont {D.}~\bibnamefont {Weygand}},\ }\href
  {\doibase 10.1146/annurev-matsci-082908-145409} {\bibfield  {journal}
  {\bibinfo  {journal} {Annual Review of Materials Research}\ }\textbf
  {\bibinfo {volume} {40}},\ \bibinfo {pages} {293} (\bibinfo {year} {2010})},\
  \Eprint
  {http://arxiv.org/abs/https://doi.org/10.1146/annurev-matsci-082908-145409}
  {https://doi.org/10.1146/annurev-matsci-082908-145409} \BibitemShut {NoStop}%
\bibitem [{\citenamefont {Sergueeva}\ \emph {et~al.}(2004)\citenamefont
  {Sergueeva}, \citenamefont {Mara}, \citenamefont {Branagan},\ and\
  \citenamefont {Mukherjee}}]{Sergueeva2004}%
  \BibitemOpen
  \bibfield  {author} {\bibinfo {author} {\bibfnamefont {A.}~\bibnamefont
  {Sergueeva}}, \bibinfo {author} {\bibfnamefont {N.}~\bibnamefont {Mara}},
  \bibinfo {author} {\bibfnamefont {D.}~\bibnamefont {Branagan}}, \ and\
  \bibinfo {author} {\bibfnamefont {A.}~\bibnamefont {Mukherjee}},\ }\href
  {\doibase 10.1016/j.scriptamat.2004.02.019} {\bibfield  {journal} {\bibinfo
  {journal} {Scripta Materialia}\ }\textbf {\bibinfo {volume} {50}},\ \bibinfo
  {pages} {1303} (\bibinfo {year} {2004})}\BibitemShut {NoStop}%
\bibitem [{\citenamefont {Ding}\ \emph {et~al.}(2015)\citenamefont {Ding},
  \citenamefont {Li}, \citenamefont {Zhang}, \citenamefont {Wu}, \citenamefont
  {Xu},\ and\ \citenamefont {Gao}}]{DING201589}%
  \BibitemOpen
  \bibfield  {author} {\bibinfo {author} {\bibfnamefont {B.}~\bibnamefont
  {Ding}}, \bibinfo {author} {\bibfnamefont {X.}~\bibnamefont {Li}}, \bibinfo
  {author} {\bibfnamefont {X.}~\bibnamefont {Zhang}}, \bibinfo {author}
  {\bibfnamefont {H.}~\bibnamefont {Wu}}, \bibinfo {author} {\bibfnamefont
  {Z.}~\bibnamefont {Xu}}, \ and\ \bibinfo {author} {\bibfnamefont
  {H.}~\bibnamefont {Gao}},\ }\href {\doibase
  https://doi.org/10.1016/j.nanoen.2015.10.002} {\bibfield  {journal} {\bibinfo
   {journal} {Nano Energy}\ }\textbf {\bibinfo {volume} {18}},\ \bibinfo
  {pages} {89 } (\bibinfo {year} {2015})}\BibitemShut {NoStop}%
\bibitem [{\citenamefont {Maa{\ss}}\ \emph {et~al.}(2015)\citenamefont
  {Maa{\ss}}, \citenamefont {Birckigt}, \citenamefont {Borchers}, \citenamefont
  {Samwer},\ and\ \citenamefont {Volkert}}]{MAA201594}%
  \BibitemOpen
  \bibfield  {author} {\bibinfo {author} {\bibfnamefont {R.}~\bibnamefont
  {Maa{\ss}}}, \bibinfo {author} {\bibfnamefont {P.}~\bibnamefont {Birckigt}},
  \bibinfo {author} {\bibfnamefont {C.}~\bibnamefont {Borchers}}, \bibinfo
  {author} {\bibfnamefont {K.}~\bibnamefont {Samwer}}, \ and\ \bibinfo {author}
  {\bibfnamefont {C.}~\bibnamefont {Volkert}},\ }\href {\doibase
  https://doi.org/10.1016/j.actamat.2015.06.062} {\bibfield  {journal}
  {\bibinfo  {journal} {Acta Materialia}\ }\textbf {\bibinfo {volume} {98}},\
  \bibinfo {pages} {94 } (\bibinfo {year} {2015})}\BibitemShut {NoStop}%
\bibitem [{\citenamefont {An}\ \emph {et~al.}(2011)\citenamefont {An},
  \citenamefont {Garrett}, \citenamefont {Samwer}, \citenamefont {Liu},
  \citenamefont {Zybin}, \citenamefont {Luo}, \citenamefont {Demetriou},
  \citenamefont {Johnson},\ and\ \citenamefont {Goddard}}]{An:2011aa}%
  \BibitemOpen
  \bibfield  {author} {\bibinfo {author} {\bibfnamefont {Q.}~\bibnamefont
  {An}}, \bibinfo {author} {\bibfnamefont {G.}~\bibnamefont {Garrett}},
  \bibinfo {author} {\bibfnamefont {K.}~\bibnamefont {Samwer}}, \bibinfo
  {author} {\bibfnamefont {Y.}~\bibnamefont {Liu}}, \bibinfo {author}
  {\bibfnamefont {S.~V.}\ \bibnamefont {Zybin}}, \bibinfo {author}
  {\bibfnamefont {S.-N.}\ \bibnamefont {Luo}}, \bibinfo {author} {\bibfnamefont
  {M.~D.}\ \bibnamefont {Demetriou}}, \bibinfo {author} {\bibfnamefont {W.~L.}\
  \bibnamefont {Johnson}}, \ and\ \bibinfo {author} {\bibfnamefont {W.~A.}\
  \bibnamefont {Goddard}},\ }\bibfield  {booktitle} {\emph {\bibinfo
  {booktitle} {The Journal of Physical Chemistry Letters}},\ }\href {\doibase
  10.1021/jz200351m} {\bibfield  {journal} {\bibinfo  {journal} {The Journal of
  Physical Chemistry Letters}\ }\textbf {\bibinfo {volume} {2}},\ \bibinfo
  {pages} {1320} (\bibinfo {year} {2011})}\BibitemShut {NoStop}%
\bibitem [{\citenamefont {Murali}\ \emph {et~al.}(2011)\citenamefont {Murali},
  \citenamefont {Guo}, \citenamefont {Zhang}, \citenamefont {Narasimhan},
  \citenamefont {Li},\ and\ \citenamefont {Gao}}]{Murali:2011aa}%
  \BibitemOpen
  \bibfield  {author} {\bibinfo {author} {\bibfnamefont {P.}~\bibnamefont
  {Murali}}, \bibinfo {author} {\bibfnamefont {T.~F.}\ \bibnamefont {Guo}},
  \bibinfo {author} {\bibfnamefont {Y.~W.}\ \bibnamefont {Zhang}}, \bibinfo
  {author} {\bibfnamefont {R.}~\bibnamefont {Narasimhan}}, \bibinfo {author}
  {\bibfnamefont {Y.}~\bibnamefont {Li}}, \ and\ \bibinfo {author}
  {\bibfnamefont {H.~J.}\ \bibnamefont {Gao}},\ }\href {\doibase
  10.1103/PhysRevLett.107.215501} {\bibfield  {journal} {\bibinfo  {journal}
  {Physical Review Letters}\ }\textbf {\bibinfo {volume} {107}},\ \bibinfo
  {pages} {215501} (\bibinfo {year} {2011})}\BibitemShut {NoStop}%
\bibitem [{\citenamefont {Guan}\ \emph {et~al.}(2013)\citenamefont {Guan},
  \citenamefont {Lu}, \citenamefont {Spector}, \citenamefont {Valavala},\ and\
  \citenamefont {Falk}}]{Guan:2013aa}%
  \BibitemOpen
  \bibfield  {author} {\bibinfo {author} {\bibfnamefont {P.}~\bibnamefont
  {Guan}}, \bibinfo {author} {\bibfnamefont {S.}~\bibnamefont {Lu}}, \bibinfo
  {author} {\bibfnamefont {M.~J.~B.}\ \bibnamefont {Spector}}, \bibinfo
  {author} {\bibfnamefont {P.~K.}\ \bibnamefont {Valavala}}, \ and\ \bibinfo
  {author} {\bibfnamefont {M.~L.}\ \bibnamefont {Falk}},\ }\href {\doibase
  10.1103/PhysRevLett.110.185502} {\bibfield  {journal} {\bibinfo  {journal}
  {Physical Review Letters}\ }\textbf {\bibinfo {volume} {110}},\ \bibinfo
  {pages} {185502} (\bibinfo {year} {2013})}\BibitemShut {NoStop}%
\bibitem [{\citenamefont {Steif}(1983)}]{STEIF1983359}%
  \BibitemOpen
  \bibfield  {author} {\bibinfo {author} {\bibfnamefont {P.~S.}\ \bibnamefont
  {Steif}},\ }\href {\doibase https://doi.org/10.1016/0022-5096(83)90005-4}
  {\bibfield  {journal} {\bibinfo  {journal} {Journal of the Mechanics and
  Physics of Solids}\ }\textbf {\bibinfo {volume} {31}},\ \bibinfo {pages} {359
  } (\bibinfo {year} {1983})}\BibitemShut {NoStop}%
\bibitem [{\citenamefont {Babu}\ \emph {et~al.}(2016)\citenamefont {Babu},
  \citenamefont {Mondal}, \citenamefont {Sengupta},\ and\ \citenamefont
  {Karmakar}}]{C5SM02200B}%
  \BibitemOpen
  \bibfield  {author} {\bibinfo {author} {\bibfnamefont {J.~S.}\ \bibnamefont
  {Babu}}, \bibinfo {author} {\bibfnamefont {C.}~\bibnamefont {Mondal}},
  \bibinfo {author} {\bibfnamefont {S.}~\bibnamefont {Sengupta}}, \ and\
  \bibinfo {author} {\bibfnamefont {S.}~\bibnamefont {Karmakar}},\ }\href
  {\doibase 10.1039/C5SM02200B} {\bibfield  {journal} {\bibinfo  {journal}
  {Soft Matter}\ }\textbf {\bibinfo {volume} {12}},\ \bibinfo {pages} {1210}
  (\bibinfo {year} {2016})}\BibitemShut {NoStop}%
\end{thebibliography}%


\begin{thebibliography}{1}%
\makeatletter
\providecommand \@ifxundefined [1]{%
 \@ifx{#1\undefined}
}%
\providecommand \@ifnum [1]{%
 \ifnum #1\expandafter \@firstoftwo
 \else \expandafter \@secondoftwo
 \fi
}%
\providecommand \@ifx [1]{%
 \ifx #1\expandafter \@firstoftwo
 \else \expandafter \@secondoftwo
 \fi
}%
\providecommand \natexlab [1]{#1}%
\providecommand \enquote  [1]{``#1''}%
\providecommand \bibnamefont  [1]{#1}%
\providecommand \bibfnamefont [1]{#1}%
\providecommand \citenamefont [1]{#1}%
\providecommand \href@noop [0]{\@secondoftwo}%
\providecommand \href [0]{\begingroup \@sanitize@url \@href}%
\providecommand \@href[1]{\@@startlink{#1}\@@href}%
\providecommand \@@href[1]{\endgroup#1\@@endlink}%
\providecommand \@sanitize@url [0]{\catcode `\\12\catcode `\$12\catcode
  `\&12\catcode `\#12\catcode `\^12\catcode `\_12\catcode `\%12\relax}%
\providecommand \@@startlink[1]{}%
\providecommand \@@endlink[0]{}%
\providecommand \url  [0]{\begingroup\@sanitize@url \@url }%
\providecommand \@url [1]{\endgroup\@href {#1}{\urlprefix }}%
\providecommand \urlprefix  [0]{URL }%
\providecommand \Eprint [0]{\href }%
\providecommand \doibase [0]{http://dx.doi.org/}%
\providecommand \selectlanguage [0]{\@gobble}%
\providecommand \bibinfo  [0]{\@secondoftwo}%
\providecommand \bibfield  [0]{\@secondoftwo}%
\providecommand \translation [1]{[#1]}%
\providecommand \BibitemOpen [0]{}%
\providecommand \bibitemStop [0]{}%
\providecommand \bibitemNoStop [0]{.\EOS\space}%
\providecommand \EOS [0]{\spacefactor3000\relax}%
\providecommand \BibitemShut  [1]{\csname bibitem#1\endcsname}%
\let\auto@bib@innerbib\@empty
\bibitem [{\citenamefont {Babu}\ \emph {et~al.}(2016)\citenamefont {Babu},
  \citenamefont {Mondal}, \citenamefont {Sengupta},\ and\ \citenamefont
  {Karmakar}}]{C5SM02200B}%
  \BibitemOpen
  \bibfield  {author} {\bibinfo {author} {\bibfnamefont {Jeetu~S.}\
  \bibnamefont {Babu}}, \bibinfo {author} {\bibfnamefont {Chandana}\
  \bibnamefont {Mondal}}, \bibinfo {author} {\bibfnamefont {Surajit}\
  \bibnamefont {Sengupta}}, \ and\ \bibinfo {author} {\bibfnamefont {Smarajit}\
  \bibnamefont {Karmakar}},\ }\bibfield  {title} {\enquote {\bibinfo {title}
  {Excess vibrational density of states and the brittle to ductile transition
  in crystalline and amorphous solids},}\ }\href {\doibase 10.1039/C5SM02200B}
  {\bibfield  {journal} {\bibinfo  {journal} {Soft Matter}\ }\textbf {\bibinfo
  {volume} {12}},\ \bibinfo {pages} {1210--1218} (\bibinfo {year}
  {2016})}\BibitemShut {NoStop}%
\end{thebibliography}%
\bibliographystyle{apsrev4-1}

\end{document}


\title{Geometry-controlled Failure Mechanisms of Amorphous Solids on the Nanoscale : Supplementary Materials}
\author{Kallol Paul$^{1}$}
\author{Ratul Dasgupta$^{2}$}
\author{J\"urgen Horbach$^{3}$}
\author{Smarajit Karmakar$^{1}$}
\affiliation{
  $^1$ TIFR Center for Interdisciplinary Science, Tata Institute of 
Fundamental Research, 36/P Gopanpally Village, Serilingampally Mandal, 
RR District, Hyderabad, 500075, Telangana, India\\
  $^2$ Department of Chemical Engineering, IIT Bombay, Powai, 
Mumbai 400076, India,\\
  $^3$ Institute f\"ur Theoretische Physik II, Heinrich-Heine-Universit\"at D\"usseldorf, 
40225 D\"usseldorf, Germany}

\maketitle

\section{Model and Simulation Details}
The model studied is a generic glass former in two dimensions. It is 
a binary mixture whose amount of bi-dispersity
was chosen to avoid crystallization as well as phase separation. A particle of type $\alpha$
and a particle of type $\beta$ ($\alpha\beta={\rm AA, AB, BB}$), separated by a distance
$r$ from each other, interact via a Lennard-Jones potential given by
\begin{equation}
\Phi_{\alpha\beta}(r) = \begin{cases}
4 \varepsilon_{\alpha\beta} \left[\left(\frac{\sigma_{\alpha \beta}}{r}\right)^{12} 
- \left(\frac{\sigma_{\alpha\beta}}{r}\right)^6\right], \quad  \text{if $r \leq r_{\rm m}\sigma_{\alpha\beta}$}\\
\varepsilon_{\alpha\beta} \left[a \left(\frac{\sigma_{\alpha\beta}}{r}\right)^{12} 
- b \left(\frac{\sigma_{\alpha\beta}}{r}\right)^6\right]\\ + 
\sum_{l=0}^{2}C_{2l}\left(\frac{r}{\sigma_{\alpha\beta}}\right)^{2l},  \quad
\text{if $ r_{\rm m}\sigma_{\alpha\beta} \leq r \leq r_{\rm c}\sigma_{\alpha\beta}$}\\
0,  \quad\quad \text{if $r \geq r_{\rm c}\sigma_{\alpha\beta}$} 
\end{cases}
\end{equation}
where \(r_{\rm m}\) is the length where the potential attains its
minimum, and \(r_{\rm c}\) is the cut-off length above which the potential
vanishes. The coefficients \(a\), \(b\) and \(C_{2l}\) are chosen
such that the repulsive and attractive parts of the potential are
continuous with two continuous derivatives at the potential minimum
and the potential goes to zero continuously at \(r_{\rm c}\) with two
continuous derivatives as well. The values of the parameters $a$, 
$b$ and $C_{2l}$ can be found in Ref.~\cite{C5SM02200B}.  

In this work, energies and lengths are given in units of $\varepsilon_{\rm
AA}$ and $\sigma_{\rm AA}$, respectively. The parameters of the
potential are $\varepsilon_{\rm AB}= 1.5 \varepsilon_{\rm AA}$,
$\varepsilon_{\rm BB}=0.5\varepsilon_{\rm AA}$, $\sigma_{\rm AB}=0.8
\sigma_{\rm AA}$, and $\sigma_{\rm BB}=0.88 \sigma_{\rm AA}$.  The
Boltzmann constant as well as mass of the particles are set to
unity, $k_{\rm B}=1.0$ and $m=1.0$, respectively. Moreover, we
set $r_{{\rm c}, \alpha\beta}=2.0\,\sigma_{\alpha\beta}$ and
$r_{{\rm m}, \alpha\beta}=2^{1/6}\,\sigma_{\alpha\beta}$.
Note that the values of $r_{{\rm c}, \alpha\beta}$ were chosen
such that at the considered density $\rho=\frac{N}{A}=1.2\,\sigma^{-2}$
(with $N$ the particle number and $A$ the area of the system) and
temperature range the system is in an intermediate state of brittle-ductile
transitions \cite{C5SM02200B}.

The equations of motion are integrated with a leap-frog algorithm, 
using an integration time step of $\delta t= 0.005\,\tau$
with $\tau=\sqrt{m \sigma_{\rm AA}^2/\varepsilon_{\rm AA}}$.
The temperature was kept constant via a Berendsen thermostat.
The number of particles \(N\) varied from 648 to 5400 and the
bi-dispersity ratio was 65 : 35 (A : B).

\section{Effect of system size}
Since $L_0$  and the density of the system is constant, increasing the aspect ratio 
basically means the system size is increased. So
one can argue that cross over in the failure mechanisms from necking to cavity 
dominated with increasing aspect ratio, is a purely a system size
effect and it is a probably not a geometric phenomenon. As we have already 
calculated the critical aspect ratio $R_C$ for a particular temperature and strain 
rate we performed a set of simulations for different system sizes with a range of
$R$ varying below and above $R_C$ and monitored the failure mechanism 
as well as the morphology of the fractured surface for each case. At least 
within the studied system sizes, it is very clear that only the aspect ratio 
controls this observed cross over in the failure mechanism with changing 
the aspect ratio. We also believe that beyond a  large system size
where the bulk effect of the system will be much more than the surface 
effect, the system will eventually behave like a brittle material failing via 
cavity formation inside the bulk at any aspect ratio of the sample. At low 
temperature and strain rate, this asymptotic bulk size is expected to be 
much larger than what we can study in a computer simulation.  

\begin{figure}[htpb]
\begin{center}
\includegraphics[scale=0.32,angle=90]{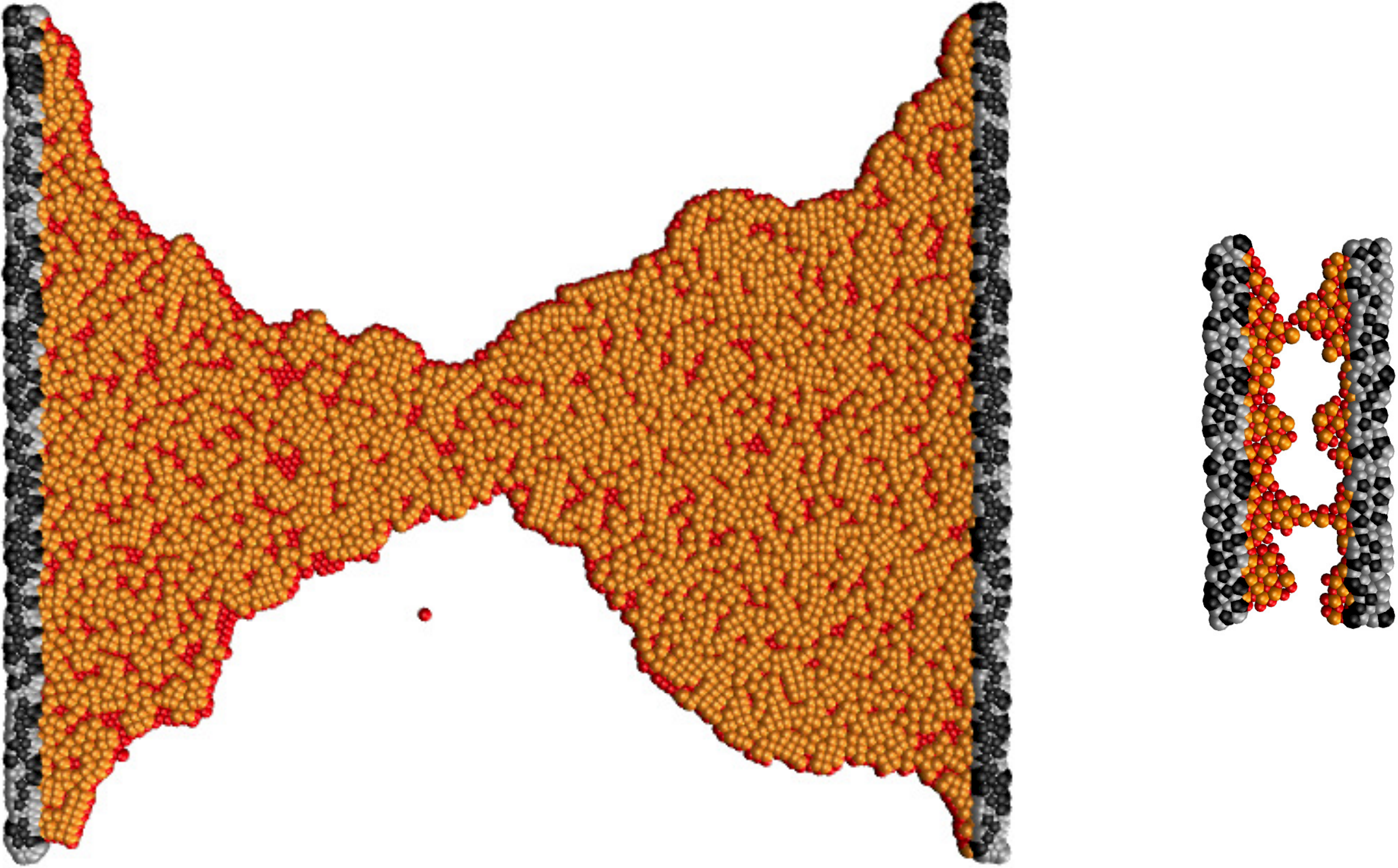}
\caption{different failure pattern: (left) \(R\) = 1.6 \(N\) = 10000, 
(right) \(R\) = 3.0 \(N\) = 500}
\label{sysDep}
\end{center}
\end{figure}

In Fig.\ref{sysDep}, we showed the 
morphology of two different cases at $T = 0.2$ and strain rate
$\dot\gamma = 10^{-5}$. In the first case, we chose a relatively large system
size $N = 10000$ and aspect ratio $R = 1.8$, which is smaller than $R_C$ at
this temperature and strain rate, whereas in the second case a relatively
very small system size $N = 500$ and a larger aspect ratio than $R_C$,
$R = 3.0$ is chosen. The morphology showed that even if in the first case, the system
size is large, the failure mechanism is like a ductile material with clear neck 
formation before the failure, whereas in the other case for a very small
system size the failure mechanism is like a brittle material with the formation
of a lot of cavities. This observation clearly confirms that at least within the 
studied system size, it is aspect ratio, not the system size that controls
the failure mechanism of these amorphous solids. 

\section{Effect of initially induced cavity}
As we have now some understanding of the existence of critical aspect ratio 
and its intimate connection to cavity formation, we tried to understand whether 
the process of cavity formation can be tuned to increase the ductility in these 
nano-scale samples. We addressed this question by studying the effect of an 
initial cavity on the failure mechanisms of the system. Our aim is to understand 
whether there is any possibility to increase the ductility of material through initially 
inducing cavities. We chose a system with a reasonably 
larger aspect ratio $R = 4.0$ where three cavities form at a 
temperature $T = 0.20$ and strain rate $\dot\gamma = 1\times10^{-5}$. 
From our earlier simulation results, the maximum ductility of this
system is found to be $\Gamma = 1.061$. Now we start with a system in 
which we remove a segment in the middle of the sample. The thickness
of the removed portion is set to twice the potential cut-off length 
$r_{c}$ to ensure zero interaction between these two segments.

\begin{figure}[htpb]
\begin{center}
\includegraphics[scale=0.32,angle=90]{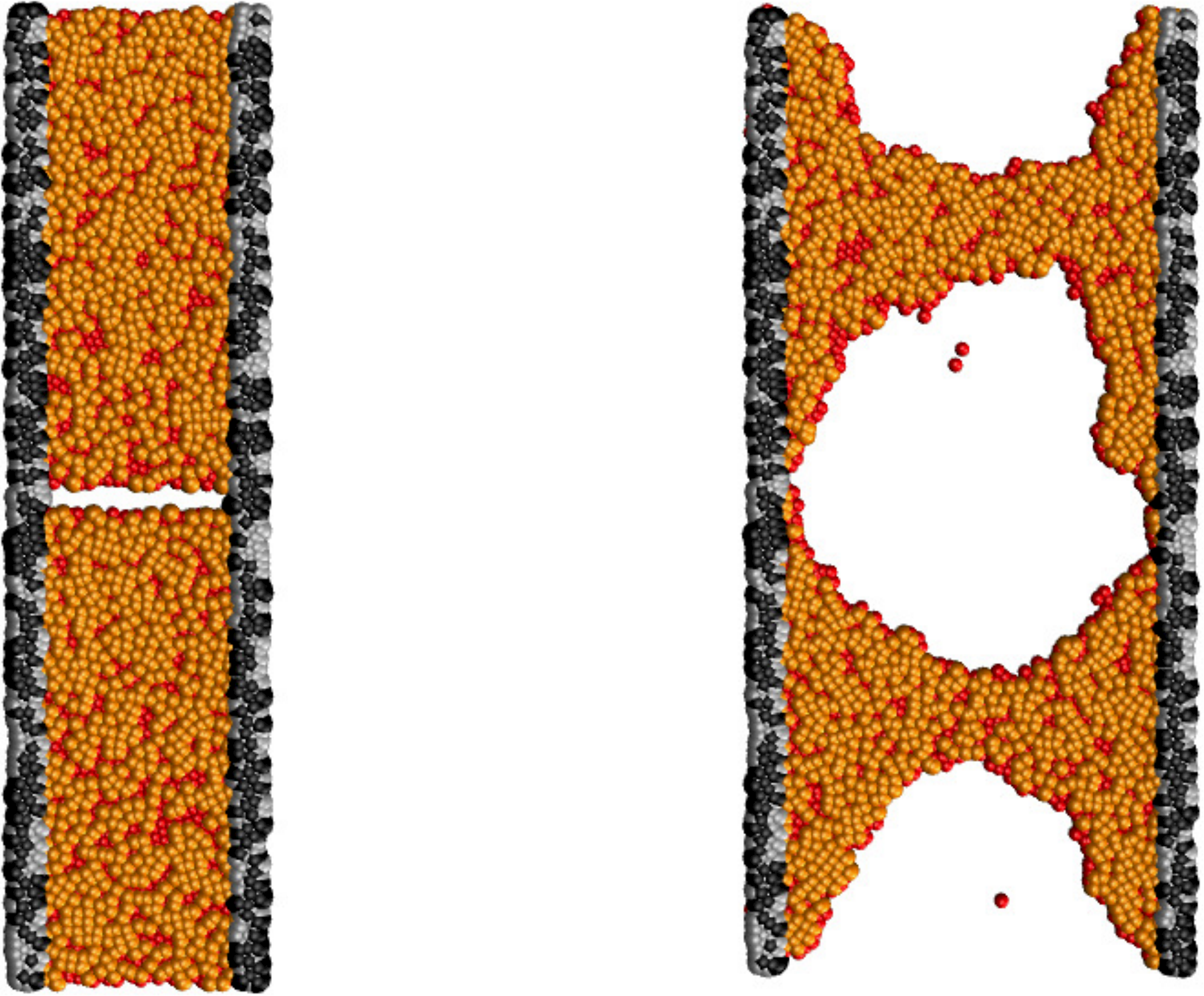}
\caption{failing pattern with initial aspect ratio \(r\) = 4.0 at 
\(T\) = 0.2 and \(\delta\) = \(10^{-5}\)}
\label{initialCavity}
\end{center}
\end{figure}

Under uniaxial elongation, we observed a much a larger cavity appearing 
in the middle instead of multiple smaller cavities with increasing load as 
shown in Fig.\ref{initialCavity}. We found that statistically,
the system was able to sustain a larger load before failure.
The measured value of $\Gamma$, in this case, is found to be
around 15\% more than the value obtained for the system with no initial 
cavity. This clearly suggests that an array of nano-pilers 
are better than bulk materials in terms of their uni-axial load
bearing capacity. This observation is again in agreement with recent
experimental results where it is suggested that an array of nano-pilers
are better candidates for supporting the load at nano-scale. 

\section{Results in three dimension}
In three dimensions, we can have a variety of geometric shapes, so to make it simpler
we have taken a bar geometry in which the length along the elongation
axis is kept at $L_0$ and cross-section of the bar is kept to be square in shape 
with area equal to $H\times H$. We then varied the cross-section area by increasing 
$H$. The aspect ratio is defined similarly as $R = H/L_0$.  

\begin{figure}[htpb]
\begin{center}
\includegraphics[scale=0.3]{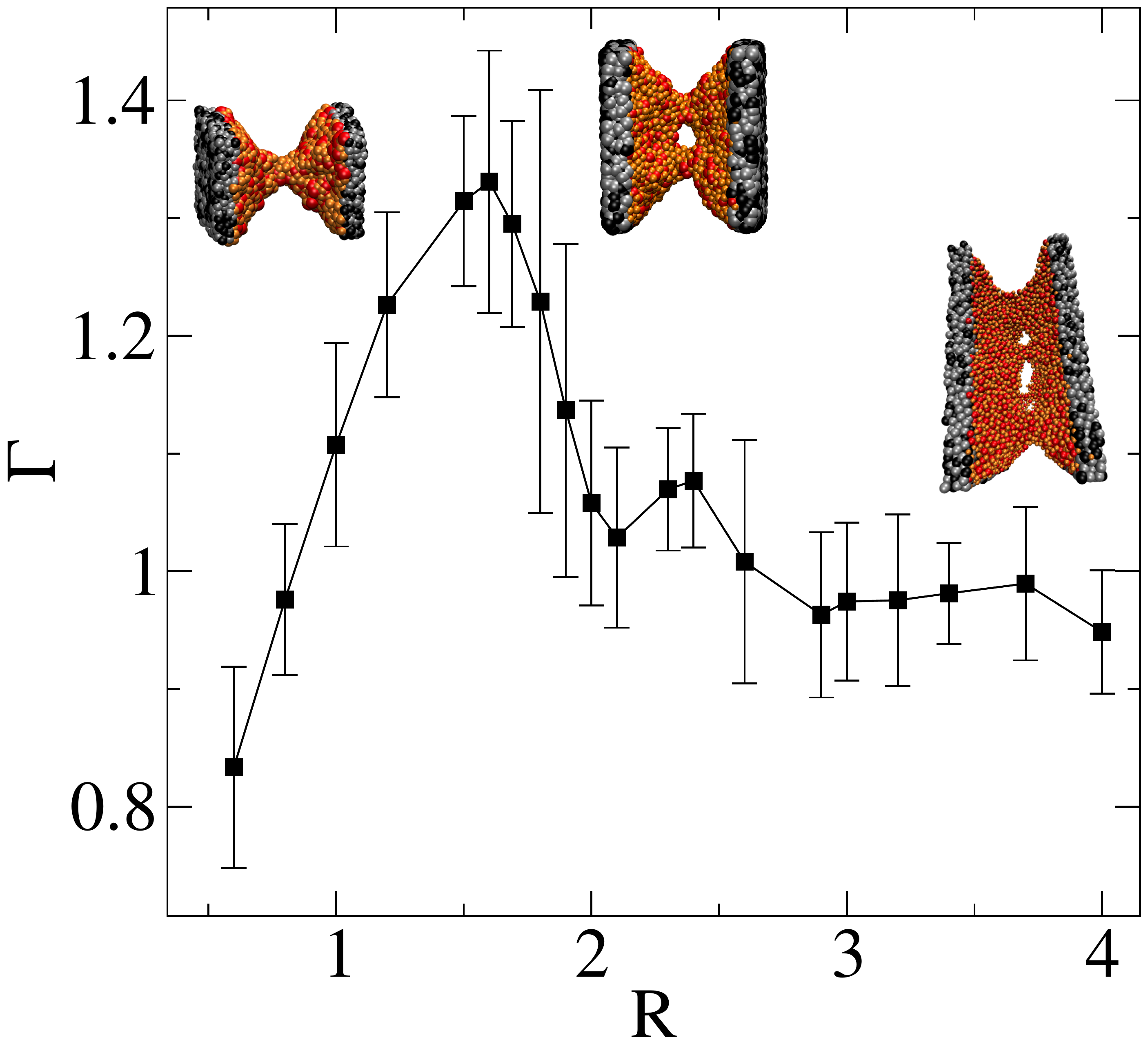}
\caption{failing pattern with initial aspect ratio \(r\) = 4.0 at 
\(T\) = 0.2 and \(\delta\) = \(10^{-5}\)}
\label{3dresults}
\end{center}
\end{figure}

In Fig.\ref{3dresults}, we have shown maximum strain $\Gamma$ for 
the three dimensional bar type sample at $T = 0.20$ and strain 
rate $\dot\gamma = 1\times 10^{-5}$. One can clearly see that maximum strain 
first increases with increasing aspect ratio before reaching a peak at a critical 
aspect ratio and then decreases quite sharply with further increase in aspect ratio. 
The results are very similar to the ones obtained for two dimensional systems. 
The failure mechanisms for systems with aspect ratio smaller than the critical 
aspect ratio is purely necking and for those with a larger aspect ratio than 
the critical aspect ratio, failure happens via cavity formation. 
For much larger aspect ratio, one observes once again the formation of multiple 
cavities and subsequent merger of them before final
failure. Thus we believe that the existence of a critical aspect ratio above 
which cavitation plays an important role in determining the failure mechanisms 
under uni-axial elongation tests are very generic and does not depend on the 
dimensionality. We believe that it will very interesting to understand the failure
mechanisms in these nano-size systems in three dimensions by
varying systematically their geometric shapes. For example, 
if the cross-sectional area is varied from square to rectangular with one 
of the side being much smaller than another side, one will
have a system which will be quasi two dimensional or like a thin
film. Failure mechanisms with increasing film thickness will be very 
important and interesting to understand for practical applications.  

\section{Curvature at a constant strain for different aspect ratio}
\begin{figure}[htpb]
\begin{center}
\vskip +0.2in
{\includegraphics[scale=0.48]{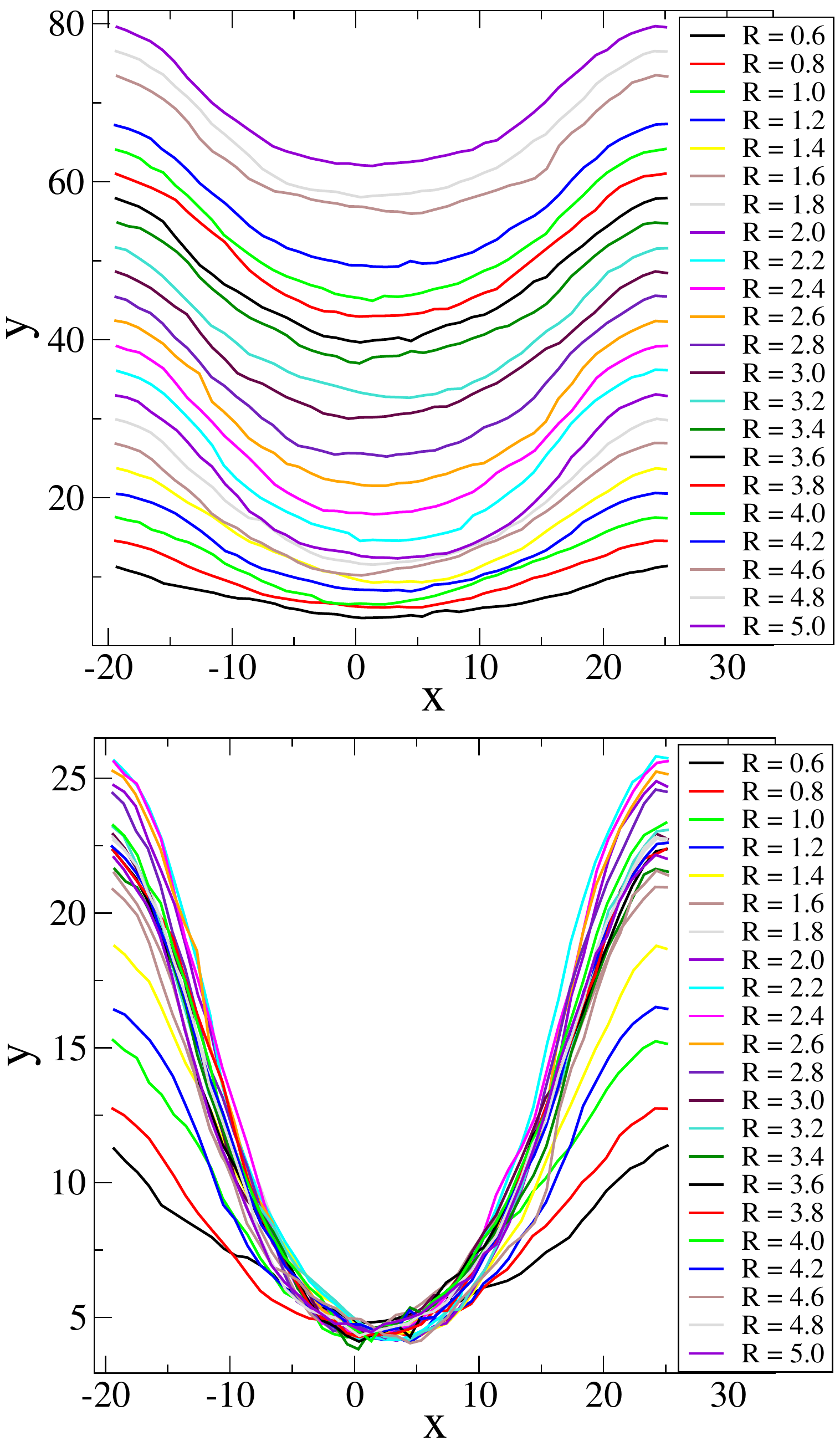}}
\caption{curvature at a constant strain for different aspect ratio}
\label{curvature}
\end{center}
\end{figure} 
We wanted to see how the curvature of free boundary bends under a 
particular strain for different aspect ratio. We chose a particular temperature 
$T = 0.20$ and strain rate $\dot{\gamma} = 5\times10^{-6}$ and monitored 
the curvature under a particular strain value $\gamma = 0.553$ for all aspect ratio.
In the top panel of Fig.\ref{curvature} we have shown the curvature of the 
free boundary for different $R$. It is clear that curvature under a fixed strain is 
increasing with $R$. In the bottom panel of Fig.\ref{curvature} we showed the 
same curvatures but putting all their middle points at the same point. Now it 
is clear that the curvature is increasing with aspect ratio up to a certain $R$ 
and then it saturates. Moreover the curvature 
 saturates as it reaches $R = 2.0$ which is exactly the critical aspect ratio at this 
 temperature and strain rate. This observation clearly says that once $R$ reaches 
 the value of corresponding $R_C$, the curvature of the free boundary immediately 
 starts to saturate and hence to create more surface it generates cavity inside the bulk.

\section{Radius of curvature as a function of strain for different aspect ratio}
\begin{figure}[htpb]
\begin{center}
{\includegraphics[scale=0.31]{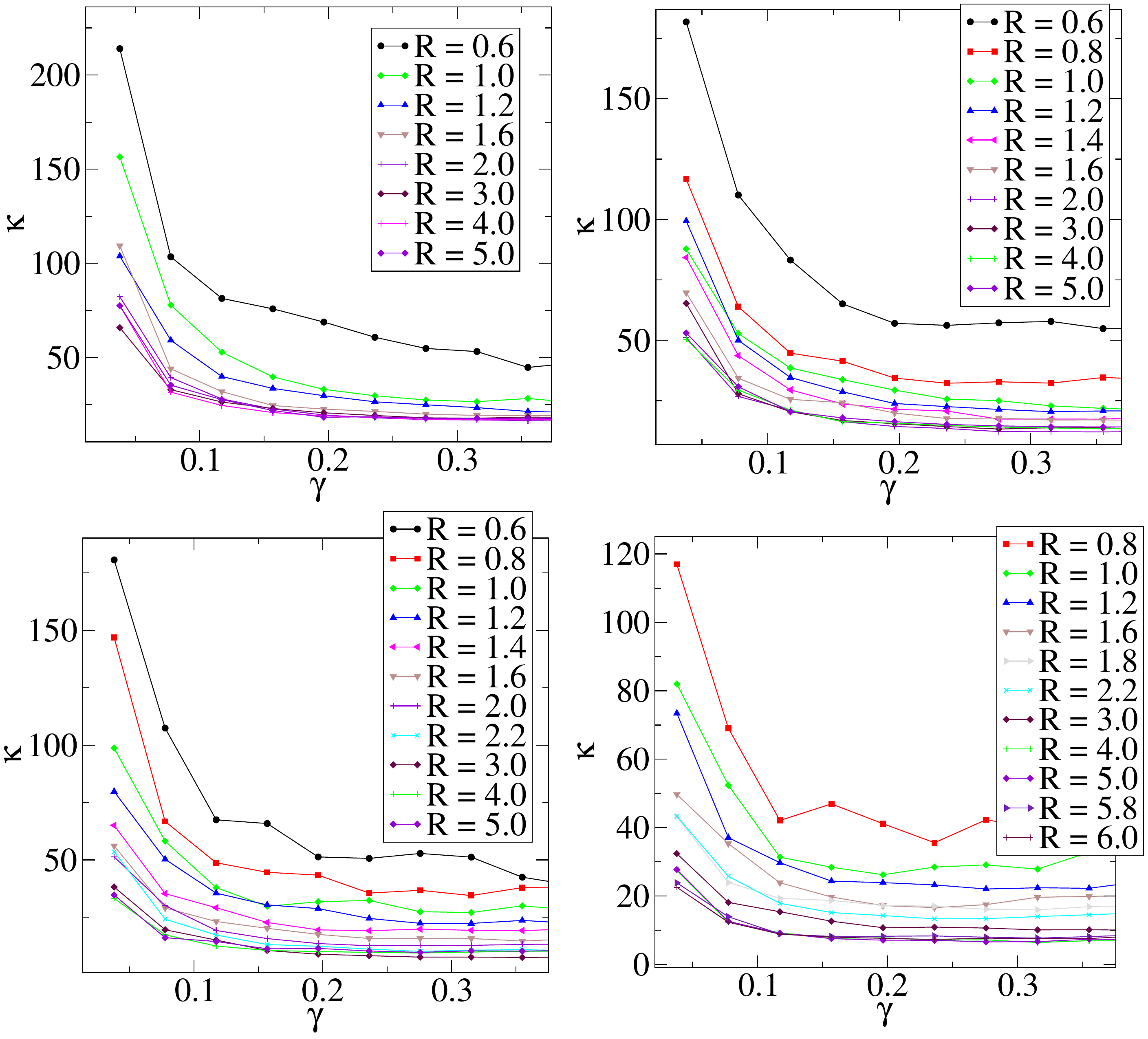}}
\caption{Radius of curvature as a function of strain at strain rate $5\times10^{-6}$ for $T = 0.15$ (top-left), for $T = 0.20$ (top-right), for $T = 0.25$ (bottom-left), for $T = 0.30$ (bottom-right)}
\label{curvatureWithStrain}
\end{center}
\end{figure}
We computed the radius of curvature ($\mathcal{K}$) as a function of strain for 
different aspect ratio. In Fig.\ref{curvatureWithStrain} we showed $\mathcal{K}$ 
as a function of strain at strain rate $\dot{\gamma} = 5\times10^{-6}$ for all 
temperatures. It is clear that for every case $\mathcal{K}$ is decreasing with 
increasing $R$ as long as $R$ reaches its critical value $R_C$. Beyond $R_C$, 
the radius of curvature saturates and it becomes the same for all aspect ratio greater 
than $R_C$. Thus the system can not generate more surface by necking and it leads 
to creating extra surface inside bulk via cavity formation. Since the critical aspect 
ratio increases with increasing temperature at a constant strain rate, 
Fig.\ref{curvatureWithStrain} clearly shows the trend of saturation of $\mathcal{K}$ 
for each temperature.

\section{Comparison of critical aspect ratio obtained via two methods} 
\begin{figure}[!tpb]
\begin{center}
{\includegraphics[scale=0.3]{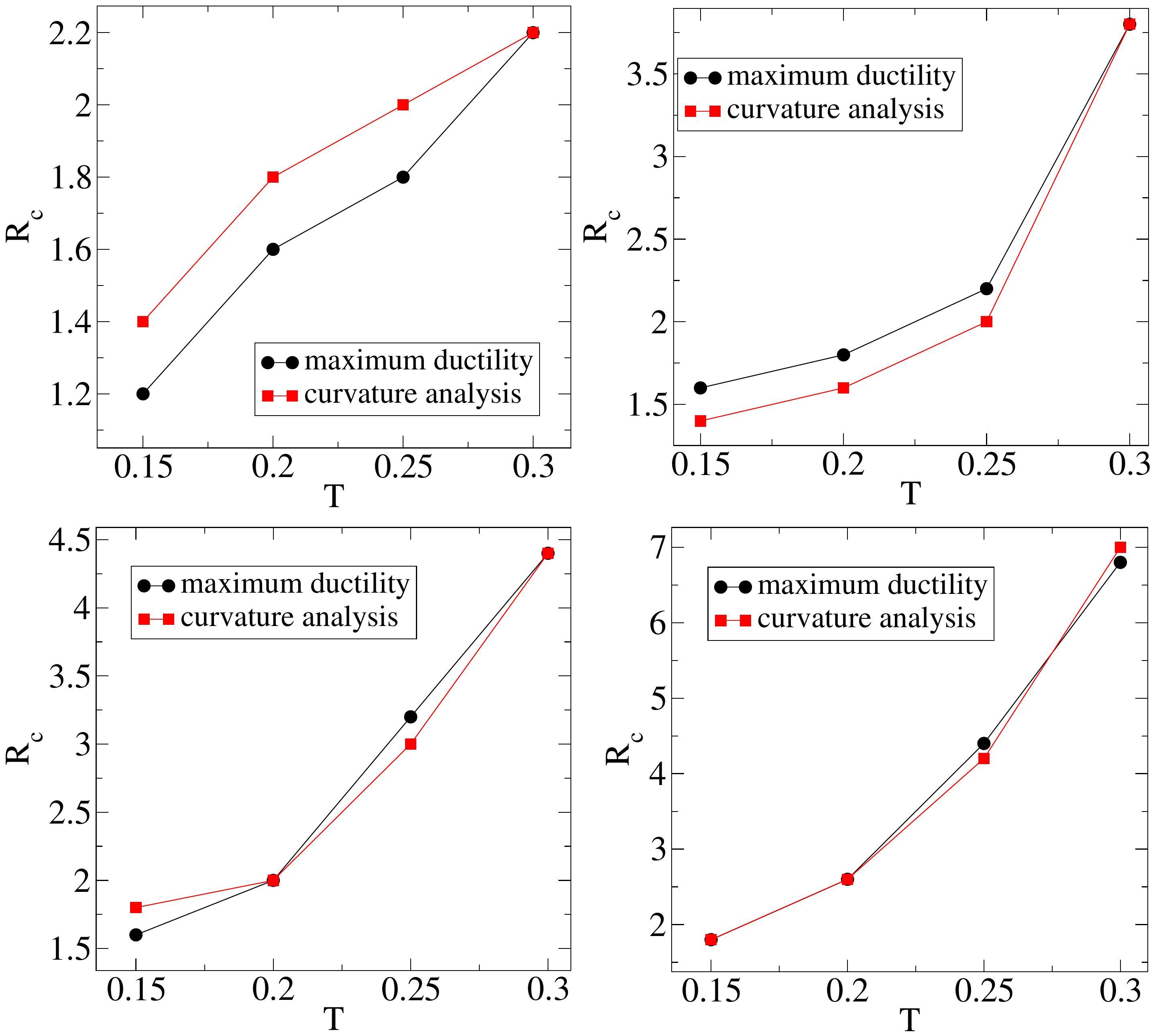}}
\caption{Comparison of critical aspect ratio between two methods for different temperature $T$ at strain rate $\dot{\gamma} = 5\times10^{-5}$ (top-left), $10^{-5}$ 
(top-right), $5\times10^{-6}$ (bottom-left), and $(10^{-6}$ 
(bottom-right)}
\label{comparison}
\end{center}
\end{figure}

Earlier we computed $R_C$ for each temperature and strain rate from the 
maximum in the ductility $\Gamma$ vs aspect ratio $R$ curve. Also using 
the new method of analysing the curvature of the free surface, we computed 
$R_C$ via eye estimation. We defined $R_C$ to be aspect ratio where the 
plot of the radius of curvature vs $R$ starts to saturate.  We then tried
to compare the value of $R_C$ that we computed from these two different 
methods in Fig.\ref{comparison}. This figure shows that our calculated value 
of $R_C$ from two different methods is very similar. It is even more closely 
if we choose higher temperature and lower strain rate.


\bibliography{siCavitation}